\documentclass[graphicx,longbibliography,reprint,prb,superscriptaddress]{revtex4-1}

\usepackage[version=3]{mhchem} % Formula subscripts using \ce{}
\usepackage[T1]{fontenc}
\usepackage{siunitx}
\usepackage{graphicx}
\usepackage{float}
\usepackage{subcaption}

\usepackage{comment}
\usepackage{color}
\usepackage{xcolor}

\newcommand{\virgolette}[1]{``#1''}
\newcommand{\ptpt}[1]{\left( #1 \right)}

\newcommand{\pgpg}[1]{\left\lbrace #1 \right\rbrace}

\newcommand{\frapar}[2]{\dfrac{\partial #1}{\partial #2}}

\newcommand{\fraparsq}[2]{\dfrac{\partial ^2 #1}{\partial #2^2}}

\usepackage[normalem]{ulem}

\useunder{\uline}{\ul}{}
\begin{document}
\author{Robin J. Dolleman}
\affiliation{Kavli Institute of Nanoscience, Delft University of Technology, Lorentzweg 1, 2628 CJ, Delft, The Netherlands}
\email{R.J.Dolleman@tudelft.nl}
\author{Pierpaolo Belardinelli}
\affiliation{Department of Precision and Microsystems Engineering, Delft University of Technology, Mekelweg 2, 2628 CD, Delft, The Netherlands}
\author{Samer Houri}
\affiliation{Kavli Institute of Nanoscience, Delft University of Technology, Lorentzweg 1, 2628 CJ, Delft, The Netherlands}
\affiliation{Current affiliation: NTT Basic Research Laboratories, NTT Corporation, 3-1, Morinosato Wakamiya, Atsugi, Kanagawa, 243-0198, Japan}
\author{Herre S.J. van der Zant}
\affiliation{Kavli Institute of Nanoscience, Delft University of Technology, Lorentzweg 1, 2628 CJ, Delft, The Netherlands}
\author{Farbod Alijani}
\affiliation{Department of Precision and Microsystems Engineering, Delft University of Technology, Mekelweg 2, 2628 CD, Delft, The Netherlands}
\author{Peter G. Steeneken}
\affiliation{Kavli Institute of Nanoscience, Delft University of Technology, Lorentzweg 1, 2628 CJ, Delft, The Netherlands}
\affiliation{Department of Precision and Microsystems Engineering, Delft University of Technology, Mekelweg 2, 2628 CD, Delft, The Netherlands}
\email{P.G.Steeneken@tudelft.nl}

\title{High-frequency stochastic switching of graphene resonators near room temperature}

\begin{abstract}
Stochastic switching between the two bistable states of a strongly driven mechanical resonator enables detection of weak signals based on probability distributions, in a manner that mimics biological systems. However, conventional silicon resonators at the microscale require a large amount of fluctuation power to achieve a switching rate in the order of a few Hertz. Here, we employ graphene membrane resonators of atomic thickness to achieve a stochastic switching rate of 7.8 kHz, which is 200 times faster than current state-of-the-art. The (effective) temperature of the fluctuations is approximately 400 K, which is 3000 times lower than the state-of-the-art. This shows that these membranes are potentially useful to transduce weak signals in the audible frequency domain. Furthermore, we perform numerical simulations to understand the transition dynamics of the resonator and derive simple analytical expressions to investigate the relevant scaling parameters that allow high-frequency, low-temperature stochastic switching to be achieved in mechanical resonators. 
\end{abstract}
\maketitle

%\section{Introduction}
Stochastic switching is the process by which a system transitions randomly between two stable states, mediated by the fluctuations in the environment. This phenomenon has been observed in a variety of physical and biological systems \cite{wiesenfeld1995stochastic,russell1999use,longtin1991time,mcnamara1988observation,hibbs1995stochastic,spano1992experimental,rouse1995flux,gammaitoni1998stochastic,lapidus1999stochastic,hales2000dynamics,tretiakov2005stochastic,wilkowski2000instabilities,ricci2017optically,rondin2017direct,levin1996broadband,douglass1993noise}. Similarly, mechanical resonators that are strongly driven can show stochastic switching between two stable attractors \cite{stambaugh2006noise,dykman1998fluctuational,chan2008paths}. This can potentially improve the transduction of small signals in a manner that mimics nature, by the stochastic resonance phenomenon \cite{badzey2005coherent,aldridge2005noise,chan2006fluctuation,ono2008noise,venstra2013stochastic}. However, high fluctuation power, far above the fluctuations present at room temperature needs to be applied to achieve stochastic switching. Despite the high resonance frequencies achieved by scaling down the resonators to the micro- or nanoscale regime, the switching rate is often quite low, in the order of 1 to 10 Hz. Extending this frequency range to the kHz regime, while lowering the fluctuation power, opens the door for new applications in the audible domain, such as ultra-sensitive microphones.  

Mechanical resonators consisting of an atomically thin membrane are ideal candidates to raise the switching rate. Their low mass ensures a MHz resonance frequency that can be easily brought in the nonlinear regime. Graphene is a single layer of carbon atoms with excellent mechanical properties \cite{novoselov2005two2,geim2007rise,lee2008measurement}. Several works have demonstrated graphene resonators \cite{bunch2007electromechanical,chen2009performance}, showing nonlinear behavior \cite{davidovikj2017nonlinear,dolleman2018opto} and several practical applications such as pressure \cite{bunch2008impermeable,smith2013electromechanical,dolleman2015graphene,eichler2011nonlinear} and gas sensors \cite{koenig2012selective,dolleman2016graphene}. The lower mass and low stiffness by virtue of the membranes thinness allows high switching rates to be achieved at lower fluctuation levels. 

Here we demonstrate high-frequency stochastic switching in strongly driven single-layer graphene drum resonators. Using an optical drive and readout, we bring the resonator into the bistable regime of the nonlinear Duffing response. By artificially adding random fluctuations to the drive, the effective temperature of the resonator is increased. We observe that the switching rate is increased with an effective temperature dependence that follows Kramer's law \cite{kramers1940brownian}. Switching rates as high as 7.8 kHz are observed close to room temperature. This work thus demonstrates a stochastic switching frequency that is more than a factor 100 higher than in prior works on mechanical resonators \cite{venstra2013stochastic}, at an effective temperature that is over a factor 3000 lower. Having a high stochastic switching rate is important to enable high-bandwidth sensing using this sensitive technique. Moreover, a low effective temperature $T_{\mathrm{eff}}$ is relevant to lower power consumption, and if $T_{\mathrm{eff}}$ can be brought down to room temperature, the intrinsic Brownian motion of the resonator can be used to enable stochastic switching based sensors. With stochastic switching frequencies above 20 Hz, this work demonstrates the potential of
graphene membranes to transduce signals in the audible frequency range.

\begin{figure*}[t!]
\includegraphics{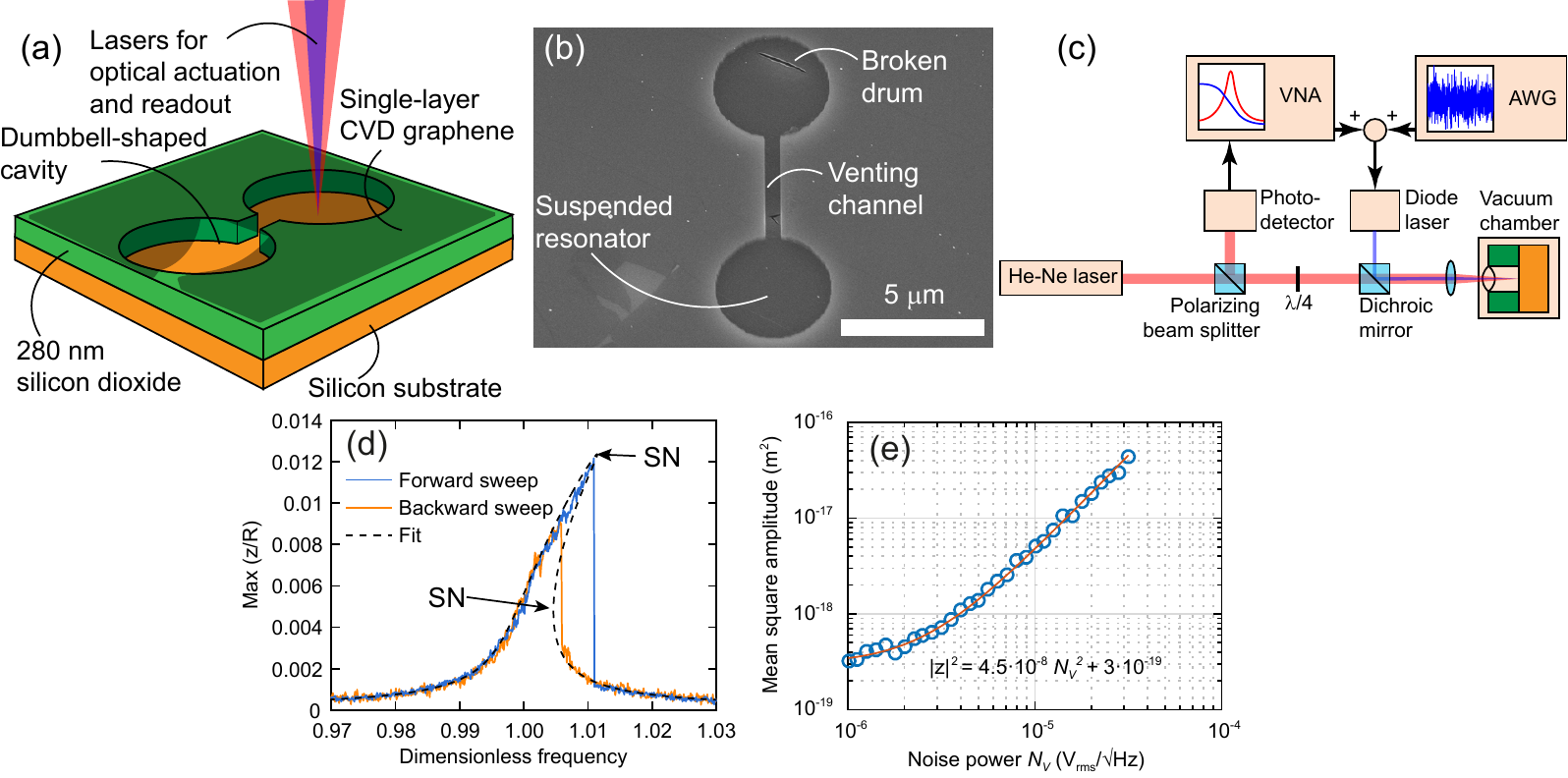}
\caption{Experimental setup. \label{fig:setupstoch} (a) Schematic figure of the sample used in the experiment. (b) Scanning electron microscope image of a successfully fabricated resonator, the top side of the dumbbell is broken and the bottom forms a resonator. (c) Laser interferometer setup used to actuate and readout the motion of the suspended graphene resonators.
(d) Frequency sweeps at high modulation power, showing the Duffing response and the bistable region. During measurements, the frequency is fixed in the center of the bistable region after finding the two saddle-node bifurcations indicated by SN in the figure. $z$ is the amplitude of the motion and $R$ is the drum radius. { The dimensionless frequency is $\Omega_F/\omega$, $\omega$ and $\Omega_F$ being the resonance frequency ($\omega = 2  \pi \times \num{13.92e6}$ rad/s) and  the drive frequency, respectively}. (e) Mean square amplitude of resonance as a function of applied noise power, this graph is used as calibration to extract the effective temperature. }
\end{figure*}
%\section{Experimental setup}
Fabrication of the samples starts with a silicon chip with a 285 nm thick thermally grown silicon dioxide layer. Dumbbell-shaped cavities as shown in Figs. \ref{fig:setupstoch}(a) and (b) are etched into the oxide layer using reactive ion etching. Single layer graphene grown by chemical vapor deposition is transferred on top of the sample using a support polymer. This polymer is dissolved and subsequently dried using critical point drying, which results in breaking of one side of the dumbbell and leaves a suspended resonator on the other end that is used for the experiment \cite{dolleman2017optomechanics}. 

%\section{Experimental setup}
Figure \ref{fig:setupstoch}(c) shows a schematic representation of the experimental setup used to actuate and detect the motion of single-layer graphene membranes. The red helium-neon laser is used to detect the motion of the membranes and the amplitude of motion is calibrated using nonlinear optical transduction \cite{dolleman2017amplitude}. The blue (405 nm) power-modulated diode laser thermally actuates the movement of the membrane, which can easily reach the bistable geometrically nonlinear regime \cite{dolleman2017optomechanics,dolleman2018opto}. A vector network analyzer (VNA, Rohde and Schwarz ZNB4-K4) actuates the membrane by sweeping the frequency forward and backward and measures the amplitude and phase of the motion. The effective temperature of the resonator is artificially raised using an arbitrary waveform generator (AWG) that outputs white noise.

 In order to quantify the effective temperature, the Brownian motion of the device is measured as a function of noise power outputted by the AWG (Fig. \ref{fig:setupstoch}(e)). From a Lorentzian fit, the mean square amplitude $\langle z^2(t) \rangle$ of the device is derived which we use to define the effective temperature $T_{\mathrm{eff}}$ \cite{hauer2013general}:
\begin{equation}
T_{\mathrm{eff}} = \frac{ m_{\mathrm{eff}} \omega^2 \langle z^2(t) \rangle}{k_B }, 
\end{equation}
 where $m_{\mathrm{eff}}$ is the modal mass, $\omega$ the resonance frequency and $k_B$ is Boltzmann's constant. The effective temperature is a means to express the fluctuation level in an intuitive manner: the fluctuations are identical to the thermal fluctuations of an undriven resonator at an actual temperature of $T=T_{\mathrm{eff}}$.

Since the amplitude is calibrated, the mean square amplitude at low fluctuation powers (where $T_{\mathrm{eff}} \approx T$, $T$ being the environmental temperature) can also be used to determine the modal mass $m_{\mathrm{eff}}$ of the resonance. From the equipartition theorem \cite{hauer2013general}:
\begin{equation}
m_{\mathrm{eff}} = \frac{k_B T}{\omega^2 \langle z^2(t) \rangle}, 
\end{equation}
we find $m_{\mathrm{eff}} =\num{1.85}$ fg. With the known modal mass, we can use the frequency response in Fig. \ref{fig:setupstoch}(d) to find the equation of motion. By fitting this frequency response we find the dimensionless equation of motion:
\begin{equation} \label{eq:motion}
\ddot{x} + 2 \zeta \dot{x} + x + \alpha x^3 = \lambda \cos{\omega_F t},
\end{equation}
with $\zeta = 0.0006$ the damping ratio, corresponding to a quality factor of 833, $\alpha = 200$ the cubic stiffness coefficient and $\lambda = \num{3e-5}$. The fundamental frequency of the resonator is 13.92 MHz.  
The equation uses the generalized coordinate $x(t)$ which represents the deflection of the membrane's center and uses scaled variables to introduce only the relevant combinations of the parameters (see Supporting Information S1).

%\section{Results}
\begin{figure*}[t!]
\includegraphics{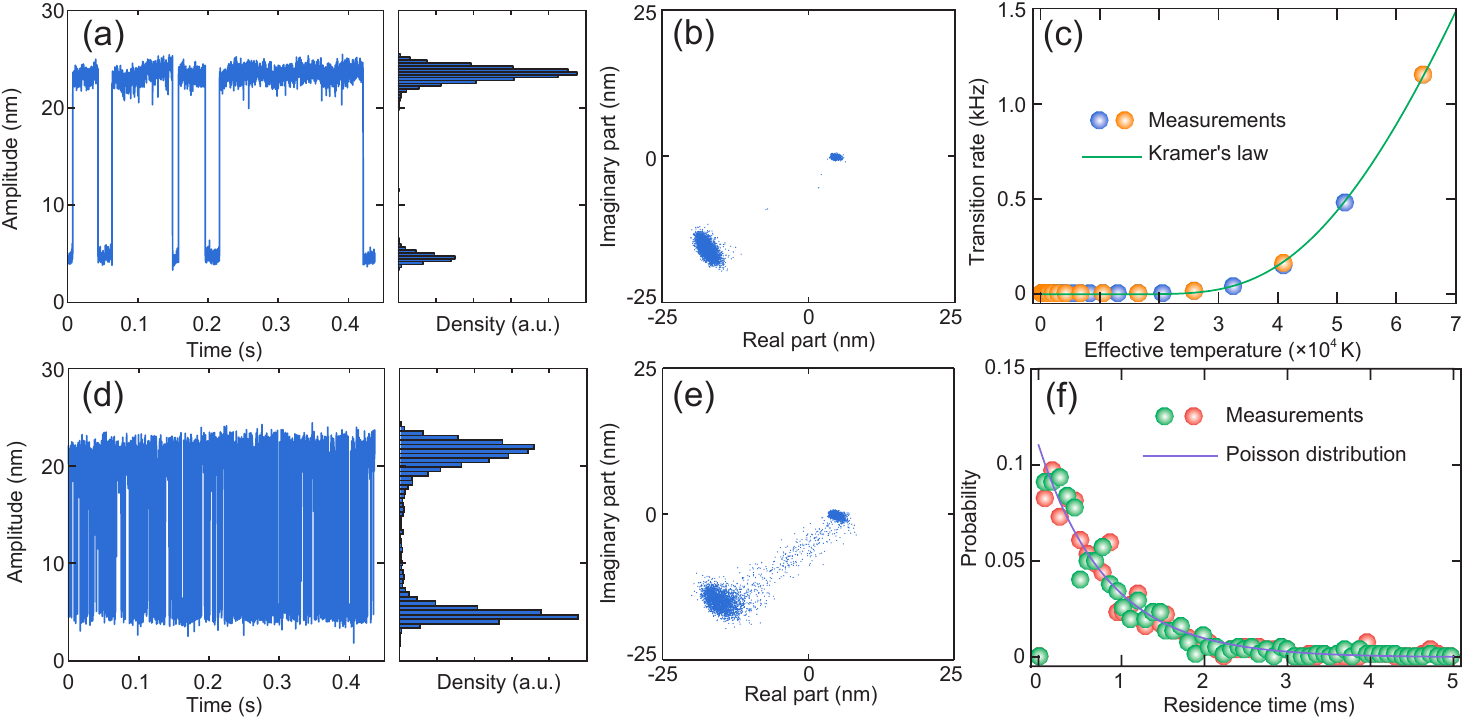}
\caption{Stochastic switching of the nonlinear resonator. a) Amplitude as function of time for an effective temperature $T_{\mathrm{eff}} = \num{25e3}$ K, showing a total of 8 fluctuation-induced transitions. b) Amplitude in the $P$-$Q$ space for  $T_{\mathrm{eff}} = \num{25e3}$ K, each point is one sample of the measurement in Fig. (a). c) Transition rate as function of effective temperature, fitted with Kramer's law (eq. \ref{eq:kramer}), two sets of consecutive measurements are shown to check for consistency. d) Amplitude as function of time for an effective temperature $T_{\mathrm{eff}} = \num{65e3}$ K, showing a total of 502 transitions. e) Amplitude in the $P$-$Q$ space for  $T_{\mathrm{eff}} = \num{65e3}$ K. f) Residence time distribution for $T_{\mathrm{eff}} = \num{65e3}$, a Poisson distribution (eq. \ref{eq:poisson}) is fitted to the data and gives a transition time $\tau_k = 0.83$ ms, corresponding to a transition rate $r_k = 1.2$ kHz.  \label{fig:3}} 
\end{figure*}
Before the experiment the resonator is prepared in a bistable state as shown in Fig. \ref{fig:setupstoch}(d). The frequency is swept forward and backward to reveal the hysteretic behavior of the device and the fixed drive frequency $\omega_F$  is then set to be in the center between the two saddle-node bifurcations.
During the experiment, the amplitude and phase of the resonator are probed as function of time using the VNA. There are now two signal sources driving the system: the fixed driving frequency from the VNA and the random fluctuations provided by the AWG.  At a fluctuation power of approximately $\num{25e3}$K the stochastic switching events are observed as shown in Fig. \ref{fig:3}(a). The amplitude $x(t)$ is split into the in-phase ($P$) and out-of-phase ($Q$) part ($x(t) = P(t) \cos{\omega_F t} + Q(t) \sin{\omega_F t}$) as shown in Fig. \ref{fig:3}(b), which reveals the two stable configurations of the resonator. Increasing the fluctuation power increases the switching rate as shown in Fig. \ref{fig:3}(d) at $\num{65e3}$K. This also causes some broadening of the stable attractors, as can be seen from Fig. \ref{fig:3}(e). The experimentally observed switching rate as function of the fluctuation power expressed in $T_{\mathrm{eff}}$ is shown in Fig. \ref{fig:3}(c). The experiment was repeated twice to check whether effects of slow frequency drift or other instabilities are affecting the experimental result, however both measurements show the same trend.  From measurements on other mechanical systems in literature, we expect the switching rate between the stable attractors to follow Kramer's law \cite{kramers1940brownian,gammaitoni1998stochastic,venstra2013stochastic,ricci2017optically}: 
\begin{equation}\label{eq:kramer}
r_k = A \exp{\left(\frac{-\Delta E}{k_B T_{\mathrm{eff}}}\right)},
\end{equation}
where $r_k$ is the transition rate, $\Delta E$ is an energy barrier, $k_B$ is the Boltzmann constant and $A$ is a parameter used for fitting. Fitting eq. \ref{eq:kramer} to the experimentally observed transition rate in Fig. \ref{fig:3}(b) shows good agreement with the experimental result. From the fit, we obtain an energy barrier of $\num{2.95}$ aJ. This energy barrier can be derived from the experimentally obtained amplitudes and effective mass as will be discussed in detail below. 

To further investigate the transition dynamics of the system, we plot the residence time distribution of two separate measurements at $\num{65e3}$K as shown in Fig. \ref{fig:3}(d). The residence time distribution should follow a Poisson distribution:
\begin{equation}\label{eq:poisson}
N(\tau) = \frac{B}{\tau_k} \exp{\left( \frac{-\tau}{\tau_k}\right)} =  B r_k \exp{\left( -\tau r_k \right)}, 
\end{equation}
which is used to fit to the experimental data.  From the fit, we find that the transition time $\tau_k = 0.83$ ms, which corresponds to a transition rate of $1.20$ kHz. This is close to the experimentally obtained value of $1.15$ kHz. 

	\begin{figure*}[t!]
	\centering
\includegraphics{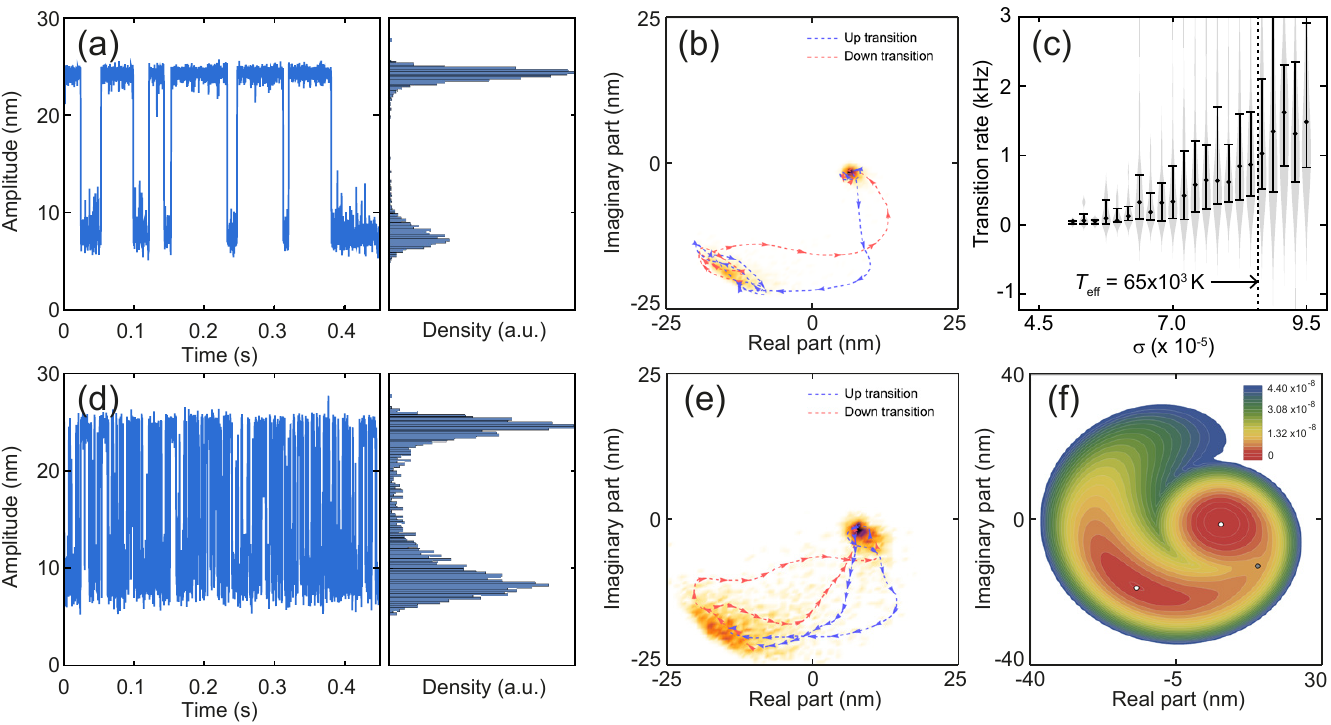}
 \caption{Simulations of stochastic switching of the nonlinear resonator in close agreement with the experiments in Fig. \ref{fig:3}. (a) Time evolution for a duration of 0.45 s of the stochastic system ($\sigma=0.000057$, $\Delta t=15$). A histogram of the distribution of the solution is shown on the right; (b) Density histogram of the solution for the long-term realization of the system. Darker regions refer to states with a more probable occurrence. (c) Distribution chart of the switching rate as a function of the imposed random fluctuations $\sigma$. The $75\%$ and $25\%$ quantile are indicated by the vertical whisker lines. (d) Time evolution of the stochastic system ($\sigma=0.000086$, $\Delta t=15$). (e) Density histogram of the solution for the long-term realization of the system. (f) Top view of the quasi-potential (see Supporting Information S2) for excitation frequency $\omega_F=1.0063$, the white dots indicate the minima while the {gray} dot indicates the saddle node. 
 }
\label{fig:sim}
\end{figure*}
%\section{Numerical simulations}
In order to further understand the dynamic behavior of the device, eq. \ref{eq:motion} is used to perform numerical simulations of the system in the presence of fluctuations to compare to the experimental results. 
We analyze the dynamics of the nonlinear oscillator using the method of averaging \cite{kryloff1947introduction, dykman1998fluctuational}. This method describes the change of the vibration amplitude in time by ironing out the fast oscillations (see Supporting Information S1 for further details). Averaging is appropriate since the quality factor is high and the transition rate is much lower than the resonance frequency.

First, a linear stability analysis is performed for the deterministic system. The eigenvalues of the linearized system predicts two stable equilibria separated by an unstable equilibrium (a saddle). 
The original model is perturbed by adding a Gaussian white noise process, with intensity $\sigma$, details of which are shown in Supporting Information S1. The intensity $\sigma$ was matched to the experiments by evaluating the mean square amplitude due to the fluctuations $\langle x^2 (t) \rangle$ from the simulations and matching them to the experimentally measured mean square amplitude in Fig. \ref{fig:setupstoch}(b). 
The stochastic switching behavior obtained via numerical integration of the stochastic differential equations can be seen in Figure \ref{fig:sim}.

We simulate a time evolution of the system as shown in Fig. \ref{fig:sim}(a), matching the time and effective temperature of the fluctuations of the experiment in Fig. \ref{fig:3}(a).  From these simulations, it can be seen that the large amplitude solution is the most probable state for the low-fluctuation configuration because the system resides for most of the time in the basin of attraction of this stable point (see the histogram in Fig. \ref{fig:sim}(a)). Fig. \ref{fig:sim}(d), which corresponds to the measurement in Fig. \ref{fig:3}(d), shows a massive number of transitions for the resonator with a more equal residence time distribution in the two separate states.
The numerical prediction is in qualitative agreement with the switching density illustrated in Figs. \ref{fig:3}(a) and (d).

%The standard linear stability analysis has only a limited utility since the fluctuations change the dynamics, and thus alter the probability of ending up in a specific steady-state solution.
The linear stability analysis of dynamical systems unveils the existence and local properties of a given steady state, but cannot provide information on { more complex systems  characterized by meta-stable attractors.}
%in which the transitions between attractors constitute a characteristic scenario.
{ Moreover, for our system in slow variables the potential function cannot be obtained by integration of the acting forces, thus it results difficult to gain further insights.}
However, the non-gradient vector field can be decomposed 
into a gradient term (quasi-potential function) and to its perpendicular constrained  remainder  (circulatory component).
While the latter causes orbits to circulate around energy level sets, the quasi potential gives a 2D surface in which
all the trajectories 
move \virgolette{downhill} in the absence of perturbations before reaching the steady states, thus satisfying Lyapunov's   global metastability condition
\cite{Zhou2012}.
The  quasi-potential function, is calculated by solving the Hamilton-Jacobi equation associated with the equations for $P$ and $Q$ \cite{Nolting2015}.
The fixed points of the deterministic skeleton of the system represent the starting point of the the standard ordered upwind method \cite{Sethian11069}. 
Then an expanding front of points is created marching the quasi-potential outward by keeping solutions at adjacent points in ascending order.
For the numerical implementation,  the 
free  R-package \textit{QPot} has been adopted\cite{Moore029777}. %Details about the quasi-potential and its computation are given in the Supporting Information S2.
%The linear stability analysis of dynamical systems unveils the existence and local properties of a given steady state, but cannot provide information on complex systems in which the transitions between attractors constitute a characteristic scenario. 
The quasi-potential gives a qualitative picture of the slow dynamics of the system, with minima near the fixed points of the system as shown in Fig. \ref{fig:sim}(f).

The probability for the membrane to undergo in large/small harmonic oscillations is related to  drive frequency $\omega_F$. Indeed, when one solution approaches the  saddle,  the 
 area of its basin of attraction  progressively shrinks whereas the other attractive set conforms as the predominant with the  deepest potential well of the system. 
The evolution of the quasi potential close to the saddle-node bifurcations of the selected bistable region is given in the Supporting Information S2.

Figs. \ref{fig:sim}(a) and (d) show broad oscillations around the low-amplitude stable equilibrium, while more confined motion is observed around the high-amplitude equilibrium state. The quasi-potential well (top-right Fig. \ref{fig:sim}(f)) associated with the low-amplitude state has a broader shape allowing for larger deviations from the equilibrium state before the transition.
The density diagrams of the solution for the long-term ($0.45$ s) realization of the system are reported in Fig.~\ref{fig:sim}(b), (c) and (e). 

At low-fluctuation levels (Fig.~\ref{fig:sim}(b)) the cloud spread is limited and the switching paths (blue and red paths in Fig. \ref{fig:sim}(b)) are concentrated in crossing the saddle ({gray} dot in Fig. \ref{fig:sim}(f)). The direction of the trajectories is in full accordance with the rotation of the orbits predicted by the stability analysis (Supporting Information S1). 
Figure \ref{fig:sim}(e) illustrates a set of paths used by the system to revert its states. Moreover, it shows a larger spread in the phase-space, due to stronger excitation of slow-dynamics around each of the fixed points, besides the higher frequency stochastic switching between low and high-amplitude states.
Finally, the switching rate as a function of the intensity of the additive Gaussian noise is reported in Fig. \ref{fig:sim}(c). For the case of $T_{\text{eff}}= \num{65e3}$ K, corresponding to $\sigma=0.000086$, the simulated transition rate is $1.05$ kHz, consistent with the experimental findings. %should I superimpose the Kramer's law here?

%\section{Discussion}
Our experiments show high-frequency stochastic switching at lower effective temperatures. It is interesting to investigate how the system can be engineered to increase the switching rate further, for example to 20 kHz for microphone applications, while reducing the temperature of the fluctuations to room temperature. To reduce the effective temperature, from eq. \ref{eq:kramer} one needs to reduce the energy barrier $\Delta E$. This energy barrier cannot be estimated from the 2D quasipotential used to understand the transition dynamics in the slow-variables. However, a simplified understanding of the dynamics can be used to estimate the value of $\Delta E$ in Kramers law that applies to a mechanical system. For this, we assume the system undergoes a harmonic motion and neglect the anharmonic part of the motion. The total mechanical energy of a system undergoing harmonic motion with amplitude $z$ and frequency $\omega$, will have a constant total mechanical energy $E$ equal to the maximum kinetic energy:
\begin{equation}
E =  \frac{1}{2} m_{\mathrm{eff}} \omega^2  z^2.
\end{equation}
Now, if a membrane undergoes harmonic oscillation in the low-amplitude attractor with amplitude $z_{\mathrm{low}}$, it will switch to the other attractor once it reaches amplitude $z_{\mathrm{saddle}}$, as it crossed the saddle (Fig. \ref{fig:sim}(f)). Thus, the work $W$ that the thermal fluctuations must do on the system to induce a switch is equal to:
\begin{equation}
W = \Delta E_{\mathrm{low}} = \frac{1}{2} m_{\mathrm{eff}} \omega^2  (z_\mathrm{saddle}^2-z_{\mathrm{low}}^2),
\end{equation}
which is the energy barrier of the low amplitude attractor. For the system oscillating in the high amplitude attractor, work must be performed by the thermal fluctuations in order to reduce the amplitude sufficiently below the saddle amplitude. This means that for the high amplitude attractor we have the energy barrier:
\begin{equation}
\Delta E_{\mathrm{high}} = \frac{1}{2} m_{\mathrm{eff}} \omega^2  (z_\mathrm{saddle}^2-z_{\mathrm{high}}^2),
\end{equation}
The switching rate from Kramers law in eq. \ref{eq:kramer} is now obtained using:
\begin{equation}
\frac{r_{\mathrm{low}}}{r_{\mathrm{high}}} \propto \mathrm{exp}{ \left(-\frac{ \Delta E_{\mathrm{low}} - \Delta E_{\mathrm{high}}}{k_B T} \right)},
\end{equation}
the energy barrier $\Delta E$ in eq. \ref{eq:kramer} is thus given by: 
\begin{equation}\label{eq:energy}
\Delta E = \frac{1}{2} m_{\mathrm{eff}} \omega^2  (z_\mathrm{high}^2-z_{\mathrm{low}}^2), 
\end{equation}
note that the energy of the saddle has dropped out, and the total energy barrier in Kramer's law for a bistable Duffing resonator is given by the difference in mechanical energy of the high and low amplitude oscillations.
Since we measured $m_{\mathrm{eff}}$ and the amplitudes (Fig. \ref{fig:3}), we can evaluate eq. \ref{eq:energy}.  We find $\Delta E = \num{3.63}$ aJ, close to the energy barrier of 2.95 aJ obtained from eq. \ref{eq:kramer}. The somewhat lower energy barrier obtained from the experimentally measured switching rate can most likely be attributed to the broadening of the attractors, induced by the fluctuations. This effectively lowers the term $ (z_\mathrm{high}^2-z_{\mathrm{low}}^2)$ in eq. \ref{eq:energy} at certain time instances. 

Equation \ref{eq:energy} gives the relevant parameters that reveal the operating parameters and the properties of the resonator desired to reduce the energy barrier in Kramer's law, to obtain low-temperature stochastic switching. For this, we assume the bistable region is close to the resonance frequency, $\omega \approx \omega_0$ and write:
\begin{equation}\label{eq:energybarrier}
\Delta E = \frac{1}{2} k_{\mathrm{eff}} (z_\mathrm{high}^2-z_{\mathrm{low}}^2).
\end{equation}
Two parameters are thus crucial to achieve stochastic switching at low temperatures. First, the effective stiffness should be as low as possible, the ultimate thinness of graphene helps to achieve this. This could be improved further by sculpting the graphene using electron or ion beams \cite{song2011atomic,fischbein2008electron}. The second parameter that can be improved is the difference between the amplitudes squared: $z_\mathrm{high}^2-z_{\mathrm{low}}^2$. This can be achieved by tuning the fixed driving frequency to be closer to the saddle node with the lowest frequency, or by driving the system at lower powers close to the critical forcing amplitude where the resonator becomes unstable. It should be noted that the prefactor $A$ in Kramer's law (Eq. \ref{eq:kramer}) has been left out of this analysis, as it depends on the curvatures of the quasi-potential function and effective damping of the slow dynamics of the system, making it more difficult to predict. However, we expect qualitatively that lowering the amplitude difference also favourably scales the pre-factor $A$, as the curvature of the quasi-potential around the equilibrium points should reduce. 

\begin{figure}[t!]
\includegraphics{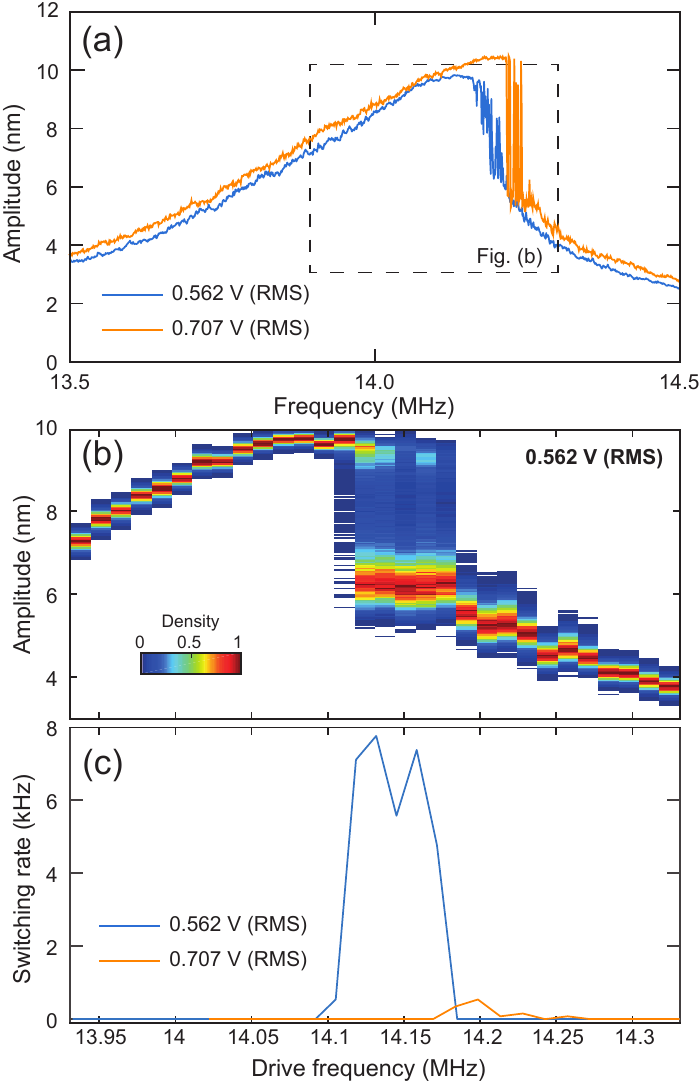}
\caption{Stochastic switching without additional noise on a different 3-micron diameter drum. a) Forward frequency sweeps at two ac driving levels, showing stochastic switching in the bistable region. b) Histogram of the amplitude at different fixed driving frequencies  at 0.562 V (RMS) driving power. c) Switching rate as function of fixed driving frequency.\label{fig:nonoise}}
\end{figure}
To qualitatively show that minimizing $z_\mathrm{high}^2-z_{\mathrm{low}}^2$ results in stochastic switching at lower temperatures, we perform an additional experiment on a different 3-micron diameter drum in Fig.~\ref{fig:nonoise}. We drive the system at two different driving levels as shown in Fig. \ref{fig:nonoise}(a), 0.562 V is almost above the critical forcing amplitude where the system becomes unstable. At these low driving levels, stochastic swtiching events are readily observed without adding noise to the system. Figure \ref{fig:nonoise}(b) shows the histogram of the amplitude at different fixed driving frequencies and Fig. \ref{fig:nonoise}(c) shows the corresponding switching rates. Close to the critical force we observe a maximum switching rate of 7762 Hz. The state-of-the-art in conventional MEMS devices achieved a 30 Hz swiching rate at an effective temperature of $\num{1.2e6}$ K \cite{venstra2013stochastic}, we have thus improved the switching rate by a factor of 200. For the effective temperature, we have to consider that the laser increases the temperature of the graphene drum somewhat. If we take the total absorbed laser power in the graphene to be roughly 0.1 mW, from measurements on similar sized drums in literature \cite{cai2010thermal} we estimate the maximum temperature in the drum to be roughly 400 K. The temperature of the fluctuations has thus been lowered by a factor of at least 3000. 

In conclusion, we have demonstrated kHz range stochastic switching on graphene drum resonators. The switching rate is two orders of magnitude higher, while the effective temperature of the fluctuations is three orders of magnitude lower than in state-of-the-art MEMS devices.
The dynamical behavior and the shape of the cycling paths are qualitatively explained by the shape of the quasi-potential around the two meta-stable equilibria that describes the system's slow dynamics. Further work can focus on increasing the switching rate and lowering of the fluctuation threshold energy $\Delta E$ to enable high-bandwidth (>10 kHz), stochastic switching enhanced, sensing at room temperature.

\begin{acknowledgements}
The authors thank Applied Nanolayers B.V. for the supply and transfer of single-layer graphene. We acknowledge Y. M. Blanter for useful discussions.
This work is part of the research programme Integrated Graphene Pressure Sensors (IGPS) with project number 13307 which is financed by the Netherlands Organisation for Scientific Research (NWO).
The research leading to these results also received funding from the European Union's Horizon 2020 research and innovation programme under grant agreement No 785219 Graphene Flagship. F.A. acknowledges support from European Research Council (ERC) grant number 802093.
\end{acknowledgements}

\newpage

\onecolumngrid

\section{Supporting information}

\section*{S1: Equations of motion}

\subsection{The deterministic skeleton}
{ The dimensionless equation that governs the dynamics of the drum is  
%%%%%%%%%%%%%%%%%%%%%%%%%%
\begin{equation}
\ddot{x}+2\zeta\dot{x}+x +\alpha x^3=\lambda \cos \omega_F t\text{.}
\label{eq_modelDimensionless}
\end{equation}
%%%%%%%%%%%%%%%%%%%%%%%%%%%
In our formulation the displacement of the membrane's center $q$ is normalized with respect to the membrane radius $R$, i.e. $x=q/R$. %The dimensionless time $t=\omega \tau$ normalizes the time variable $\tau$ by making use of the resonant frequency $\omega$.
The time variable $\tau$ is made dimensionless by making use of the resonant frequency $\omega$. The overdot in eq.~\eqref{eq_modelDimensionless} means differentiation with respect to the dimensionless time $t=\omega \tau$.
The amplitude and frequency  of the excitation are $f$ and $\Omega_F$ , related to their dimensionless counterparts $\lambda=f/\ptpt{R {\omega}^2 m_{\mathrm{eff}}}$ and $\omega_F=\Omega_F/\omega$, respectively. The symbol $m_{\mathrm{eff}}$ indicates the effective mass of the drum.
The membrane damping $c$ is  scaled to the  dimensionless damping ratio  $2 \zeta= c/\ptpt{\omega m_{\mathrm{eff}}}$. Finally, $\alpha=R^2 k_3/\ptpt{{\omega}^2 m_{\mathrm{eff}}}$ is the dimensionless conjugate of the cubic stiffness coeficient $k_3$.

In order to analyze the slow dynamical evolution of the system, the solution is assumed to have the form
}
%%%%%%%%%%%%%%%%%%%%%%%%%%
\begin{equation}
x(t)=P\ptpt{t} \cos \omega_F t + Q\ptpt{t} \sin \omega_F t,
\label{eq_solution}
\end{equation}
%%%%%%%%%%%%%%%%%%%%%%%%%%%
in which $P\ptpt{t}$ and $Q\ptpt{t}$ are slowly varying functions of time. Following the method of variation of parameters, the solution is subject to the condition \cite{kryloff1947introduction}:
%%%%%%%%%%%%%%%%%%%%%%%%%%
\begin{equation}
\dot{P}\ptpt{t} \cos \omega_F t + \dot{Q}\ptpt{t} \sin \omega_F t=0 \text{.}
\label{eq_constrain}
\end{equation}
%%%%%%%%%%%%%%%%%%%%%%%%%%%
By substituting eq. \ref{eq_solution} with its corresponding time derivatives into eq. \ref{eq_modelDimensionless}, and making use of eq. \ref{eq_constrain}, we obtain:
%%%%%%%%%%%%%%%%%%%%%%%%%%
\begin{equation}
\left\{
\begin{array}{l}
\displaystyle \dot{P}=\dfrac{{\omega}^2-{\omega_F}^2}{2 \omega_F}Q-\Gamma P+\dfrac{3}{8}\dfrac{\gamma}{\omega_F} Q\ptpt{P^2+Q^2}
\\
\displaystyle \dot{Q}=-\dfrac{{\omega}^2-{\omega_F}^2}{2 \omega_F}P-\Gamma Q-\dfrac{3}{8}\dfrac{\gamma}{\omega_F} P\ptpt{P^2+Q^2}+\dfrac{F}{2 \omega_F}
\text{.}
\end{array}\right.
\label{eq_modelsystem}
\end{equation}
%%%%%%%%%%%%%%%%%%%%%%%%%%%
\begin{figure}%[H]
	\centering
	\includegraphics[width=0.4\textwidth]{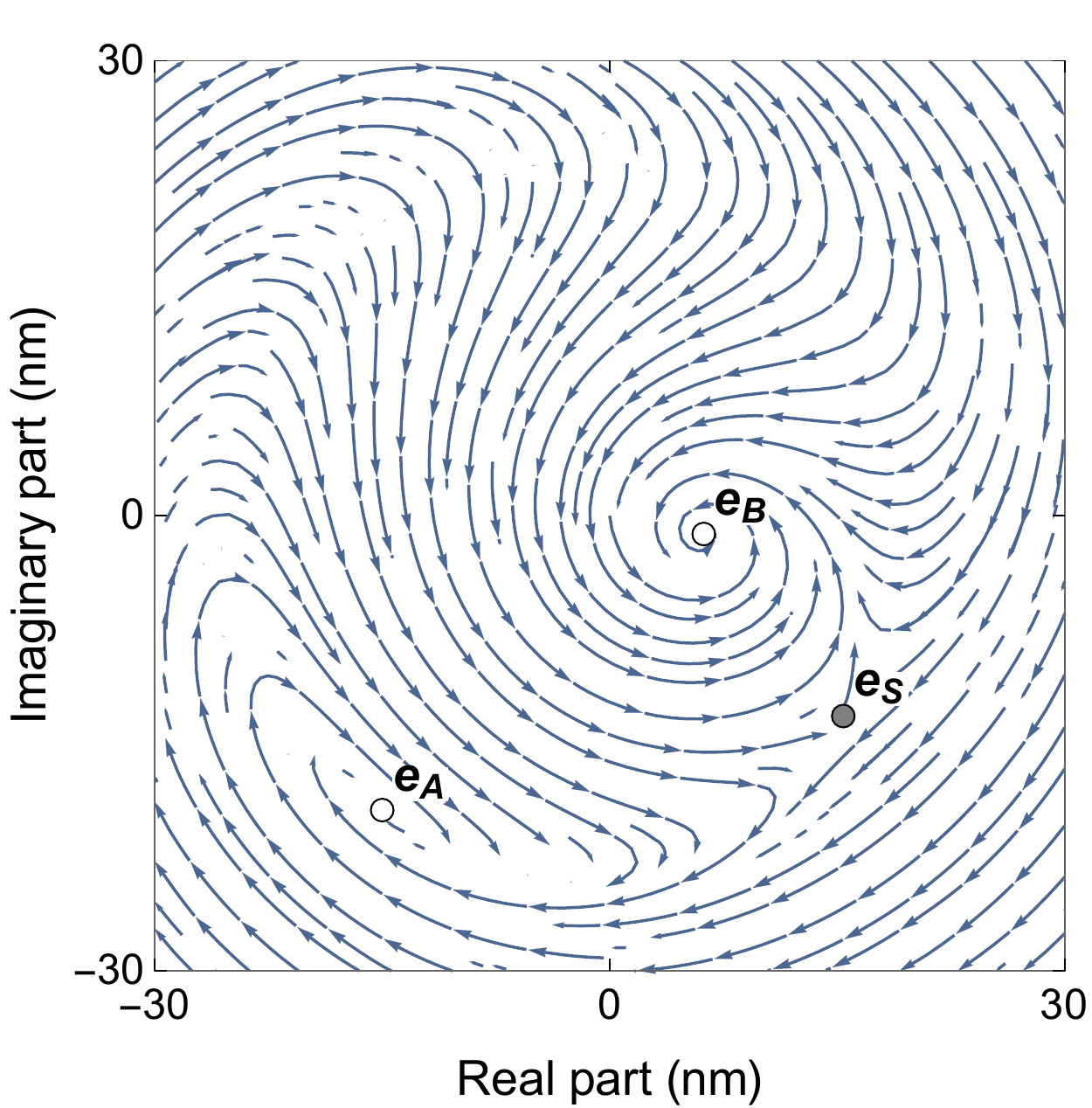}
 \caption{Stream plot for the deterministic vector field of the  model in eq.~\eqref{eq_modelsystem}. Unstable and stable equilibrium points are reported with  white and gray disks, respectively. and stable equilibria are gray disks.  Blue lines and arrows show the direction of trajectories $\omega_F=1.0063$. 
 }
\label{streamplot}
\end{figure}

\subsection{Stochastic differential system}

The deterministic skeleton of the system (Fig.~\ref{streamplot}) shows 3 equilibria:
$e_A=\pgpg{0.00246073, -0.000508264}$, 
  $e_B=\pgpg{-0.00600831, -0.00778412 }$ and
  $e_S=\pgpg{ 0.00613994, -0.00527489}$.
A linear stability analysis tells us that $e_A$ and $e_B$ are stable equilibrium, whereas $e_S$ is a saddle point. The real part of the eigenvalues of the Jacobian 
for the stable equilibrium is the same for both the stable equilibrium points ($-0.00012 \pm 0.00376934 i$ for $e_A$, $-0.00012 \pm 0.0053184 i$ for $e_B$, $-0.00461927$ and $ 0.00221927$ for $e_S$)  suggesting an equal stability.

The  deterministic  system of eqs.~\eqref{eq_modelsystem} is then perturbed by a Gaussian white noise process with intensity $\sigma_1$ and $\sigma_2$ in the equations for $\dot{P}$ and $\dot{Q}$, respectively. 
The system of stochastic differential equations (SDE) with additive noise is:
%%%%%%%%%%%%%%%%%%%%%%%%%%
\begin{equation}
\left\{
\begin{array}{l}
\displaystyle dP=\ptpt{\dfrac{{\omega}^2-{\omega_F}^2}{2 \omega_F}Q-\Gamma P +\dfrac{3}{8}\dfrac{\gamma}{\omega_F} Q\ptpt{P^2+Q^2}}dt +\sigma_1 d W_1
\\
\displaystyle dQ=\ptpt{-\dfrac{{\omega}^2-{\omega_F}^2}{2 \omega_F}P-\Gamma Q-\dfrac{3}{8}\dfrac{\gamma}{\omega_F} P\ptpt{P^2+Q^2}+\dfrac{F}{2 \omega_F}}dt+\sigma_2 d W_2
\end{array}\right.
\label{eq_modelSystemstochastic}
\end{equation}
%%%%%%%%%%%%%%%%%%%%%%%%%%%
in which $W_1(t)$ and $W_2(t)$ are independent Wiener processes, normally distributed random variables with mean zero and variance \textit{dt}.
Note that neither $W$ nor the state variables $P$ and $Q$ are anywhere differentiable now that the system is converted to a set of stochastic differential equations. For the integration of eq.~\ref{eq_modelSystemstochastic}, the It\^{o} scheme will be employed \cite{oksendal2013stochastic}. 

{
\section*{S2: Quasi-potential evolution in the bistable region}
%\subsection{Effect of the excitation frequency}

Here we report the conformation of the quasi potential for different values of the excitation $\omega_F$ in Eq.~\eqref{qpvariation}. 
\begin{figure}%[H]
	\centering
	\includegraphics[width=1.0\textwidth]{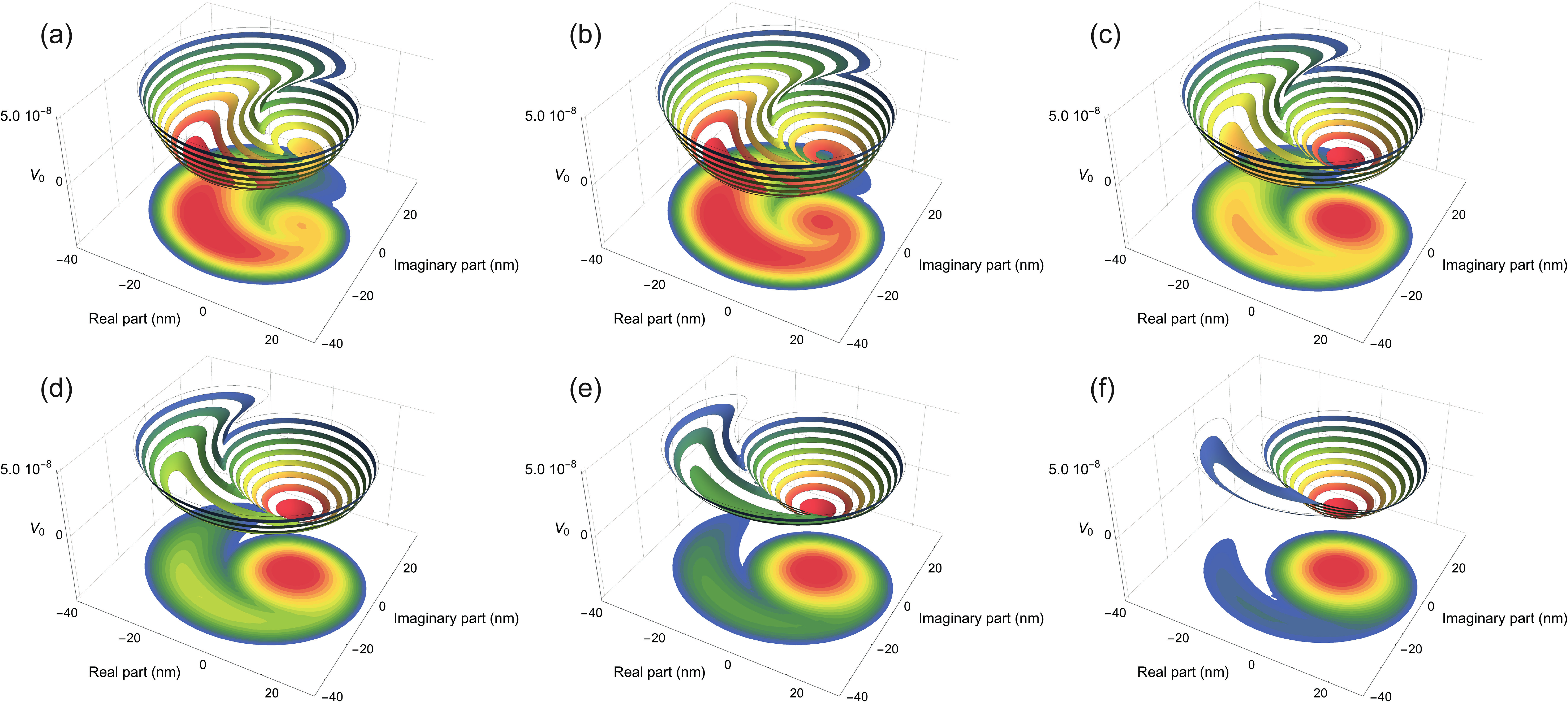}
 \caption{The quasi-potential function for different excitation frequencies.
 a) $\omega_F=1.005$;
 b) $\omega_F=1.006$;
 c) $\omega_F=1.007$;
 d) $\omega_F=1.008$;
 e) $\omega_F=1.009$;
 f) $\omega_F=1.010$.
 }
\label{qpvariation}
\end{figure}
The shape of the quasi-potential surface changes drastically with the drive frequency.
In Fig.~\ref{qpvariation} we show the change of the system  from one meta-stable equilibrium to the other.
Approaching the membrane resonance through large frequencies ($\omega_F=1.0010$), the 
low amplitude solution loses progressively its stability increasing the extension of the basin for the high amplitude resonant solution (see Fig.~\ref{qpvariation}(a)). 
The analysis presented in the main manuscript for $\omega_F=1.0063$ realizes an intermediate situation with wells not dissimilar in depth. However, thanks to the quasi-potential, we observe that the well for the low-amplitude solution is more confined and able to trap orbits.
}


\begin{thebibliography}{50}%
\makeatletter
\providecommand \@ifxundefined [1]{%
 \@ifx{#1\undefined}
}%
\providecommand \@ifnum [1]{%
 \ifnum #1\expandafter \@firstoftwo
 \else \expandafter \@secondoftwo
 \fi
}%
\providecommand \@ifx [1]{%
 \ifx #1\expandafter \@firstoftwo
 \else \expandafter \@secondoftwo
 \fi
}%
\providecommand \natexlab [1]{#1}%
\providecommand \enquote  [1]{``#1''}%
\providecommand \bibnamefont  [1]{#1}%
\providecommand \bibfnamefont [1]{#1}%
\providecommand \citenamefont [1]{#1}%
\providecommand \href@noop [0]{\@secondoftwo}%
\providecommand \href [0]{\begingroup \@sanitize@url \@href}%
\providecommand \@href[1]{\@@startlink{#1}\@@href}%
\providecommand \@@href[1]{\endgroup#1\@@endlink}%
\providecommand \@sanitize@url [0]{\catcode `\\12\catcode `\$12\catcode
  `\&12\catcode `\#12\catcode `\^12\catcode `\_12\catcode `\%12\relax}%
\providecommand \@@startlink[1]{}%
\providecommand \@@endlink[0]{}%
\providecommand \url  [0]{\begingroup\@sanitize@url \@url }%
\providecommand \@url [1]{\endgroup\@href {#1}{\urlprefix }}%
\providecommand \urlprefix  [0]{URL }%
\providecommand \Eprint [0]{\href }%
\providecommand \doibase [0]{http://dx.doi.org/}%
\providecommand \selectlanguage [0]{\@gobble}%
\providecommand \bibinfo  [0]{\@secondoftwo}%
\providecommand \bibfield  [0]{\@secondoftwo}%
\providecommand \translation [1]{[#1]}%
\providecommand \BibitemOpen [0]{}%
\providecommand \bibitemStop [0]{}%
\providecommand \bibitemNoStop [0]{.\EOS\space}%
\providecommand \EOS [0]{\spacefactor3000\relax}%
\providecommand \BibitemShut  [1]{\csname bibitem#1\endcsname}%
\let\auto@bib@innerbib\@empty
%</preamble>
\bibitem [{\citenamefont {Wiesenfeld}\ and\ \citenamefont
  {Moss}(1995)}]{wiesenfeld1995stochastic}%
  \BibitemOpen
  \bibfield  {author} {\bibinfo {author} {\bibfnamefont {Kurt}\ \bibnamefont
  {Wiesenfeld}}\ and\ \bibinfo {author} {\bibfnamefont {Frank}\ \bibnamefont
  {Moss}},\ }\bibfield  {title} {\enquote {\bibinfo {title} {Stochastic
  resonance and the benefits of noise: from ice ages to crayfish and squids},}\
  }\href@noop {} {\bibfield  {journal} {\bibinfo  {journal} {Nature}\ }\textbf
  {\bibinfo {volume} {373}},\ \bibinfo {pages} {33} (\bibinfo {year}
  {1995})}\BibitemShut {NoStop}%
\bibitem [{\citenamefont {Russell}\ \emph {et~al.}(1999)\citenamefont
  {Russell}, \citenamefont {Wilkens},\ and\ \citenamefont
  {Moss}}]{russell1999use}%
  \BibitemOpen
  \bibfield  {author} {\bibinfo {author} {\bibfnamefont {David~F}\ \bibnamefont
  {Russell}}, \bibinfo {author} {\bibfnamefont {Lon~A}\ \bibnamefont
  {Wilkens}}, \ and\ \bibinfo {author} {\bibfnamefont {Frank}\ \bibnamefont
  {Moss}},\ }\bibfield  {title} {\enquote {\bibinfo {title} {Use of behavioural
  stochastic resonance by paddle fish for feeding},}\ }\href@noop {} {\bibfield
   {journal} {\bibinfo  {journal} {Nature}\ }\textbf {\bibinfo {volume}
  {402}},\ \bibinfo {pages} {291} (\bibinfo {year} {1999})}\BibitemShut
  {NoStop}%
\bibitem [{\citenamefont {Longtin}\ \emph {et~al.}(1991)\citenamefont
  {Longtin}, \citenamefont {Bulsara},\ and\ \citenamefont
  {Moss}}]{longtin1991time}%
  \BibitemOpen
  \bibfield  {author} {\bibinfo {author} {\bibfnamefont {Andr{\'e}}\
  \bibnamefont {Longtin}}, \bibinfo {author} {\bibfnamefont {Adi}\ \bibnamefont
  {Bulsara}}, \ and\ \bibinfo {author} {\bibfnamefont {Frank}\ \bibnamefont
  {Moss}},\ }\bibfield  {title} {\enquote {\bibinfo {title} {Time-interval
  sequences in bistable systems and the noise-induced transmission of
  information by sensory neurons},}\ }\href@noop {} {\bibfield  {journal}
  {\bibinfo  {journal} {Physical Review Letters}\ }\textbf {\bibinfo {volume}
  {67}},\ \bibinfo {pages} {656} (\bibinfo {year} {1991})}\BibitemShut
  {NoStop}%
\bibitem [{\citenamefont {McNamara}\ \emph {et~al.}(1988)\citenamefont
  {McNamara}, \citenamefont {Wiesenfeld},\ and\ \citenamefont
  {Roy}}]{mcnamara1988observation}%
  \BibitemOpen
  \bibfield  {author} {\bibinfo {author} {\bibfnamefont {Bruce}\ \bibnamefont
  {McNamara}}, \bibinfo {author} {\bibfnamefont {Kurt}\ \bibnamefont
  {Wiesenfeld}}, \ and\ \bibinfo {author} {\bibfnamefont {Rajarshi}\
  \bibnamefont {Roy}},\ }\bibfield  {title} {\enquote {\bibinfo {title}
  {Observation of stochastic resonance in a ring laser},}\ }\href@noop {}
  {\bibfield  {journal} {\bibinfo  {journal} {Physical Review Letters}\
  }\textbf {\bibinfo {volume} {60}},\ \bibinfo {pages} {2626} (\bibinfo {year}
  {1988})}\BibitemShut {NoStop}%
\bibitem [{\citenamefont {Hibbs}\ \emph {et~al.}(1995)\citenamefont {Hibbs},
  \citenamefont {Singsaas}, \citenamefont {Jacobs}, \citenamefont {Bulsara},
  \citenamefont {Bekkedahl},\ and\ \citenamefont {Moss}}]{hibbs1995stochastic}%
  \BibitemOpen
  \bibfield  {author} {\bibinfo {author} {\bibfnamefont {AD}~\bibnamefont
  {Hibbs}}, \bibinfo {author} {\bibfnamefont {AL}~\bibnamefont {Singsaas}},
  \bibinfo {author} {\bibfnamefont {EW}~\bibnamefont {Jacobs}}, \bibinfo
  {author} {\bibfnamefont {AR}~\bibnamefont {Bulsara}}, \bibinfo {author}
  {\bibfnamefont {JJ}~\bibnamefont {Bekkedahl}}, \ and\ \bibinfo {author}
  {\bibfnamefont {F}~\bibnamefont {Moss}},\ }\bibfield  {title} {\enquote
  {\bibinfo {title} {Stochastic resonance in a superconducting loop with a
  {Josephson} junction},}\ }\href@noop {} {\bibfield  {journal} {\bibinfo
  {journal} {Journal of Applied Physics}\ }\textbf {\bibinfo {volume} {77}},\
  \bibinfo {pages} {2582--2590} (\bibinfo {year} {1995})}\BibitemShut {NoStop}%
\bibitem [{\citenamefont {Spano}\ \emph {et~al.}(1992)\citenamefont {Spano},
  \citenamefont {Wun-Fogle},\ and\ \citenamefont
  {Ditto}}]{spano1992experimental}%
  \BibitemOpen
  \bibfield  {author} {\bibinfo {author} {\bibfnamefont {ML}~\bibnamefont
  {Spano}}, \bibinfo {author} {\bibfnamefont {M}~\bibnamefont {Wun-Fogle}}, \
  and\ \bibinfo {author} {\bibfnamefont {WL}~\bibnamefont {Ditto}},\ }\bibfield
   {title} {\enquote {\bibinfo {title} {Experimental observation of stochastic
  resonance in a magnetoelastic ribbon},}\ }\href@noop {} {\bibfield  {journal}
  {\bibinfo  {journal} {Physical Review A}\ }\textbf {\bibinfo {volume} {46}},\
  \bibinfo {pages} {5253} (\bibinfo {year} {1992})}\BibitemShut {NoStop}%
\bibitem [{\citenamefont {Rouse}\ \emph {et~al.}(1995)\citenamefont {Rouse},
  \citenamefont {Han},\ and\ \citenamefont {Lukens}}]{rouse1995flux}%
  \BibitemOpen
  \bibfield  {author} {\bibinfo {author} {\bibfnamefont {R}~\bibnamefont
  {Rouse}}, \bibinfo {author} {\bibfnamefont {Siyuan}\ \bibnamefont {Han}}, \
  and\ \bibinfo {author} {\bibfnamefont {JE}~\bibnamefont {Lukens}},\
  }\bibfield  {title} {\enquote {\bibinfo {title} {Flux amplification using
  stochastic superconducting quantum interference devices},}\ }\href@noop {}
  {\bibfield  {journal} {\bibinfo  {journal} {Applied Physics Letters}\
  }\textbf {\bibinfo {volume} {66}},\ \bibinfo {pages} {108--110} (\bibinfo
  {year} {1995})}\BibitemShut {NoStop}%
\bibitem [{\citenamefont {Gammaitoni}\ \emph {et~al.}(1998)\citenamefont
  {Gammaitoni}, \citenamefont {H{\"a}nggi}, \citenamefont {Jung},\ and\
  \citenamefont {Marchesoni}}]{gammaitoni1998stochastic}%
  \BibitemOpen
  \bibfield  {author} {\bibinfo {author} {\bibfnamefont {Luca}\ \bibnamefont
  {Gammaitoni}}, \bibinfo {author} {\bibfnamefont {Peter}\ \bibnamefont
  {H{\"a}nggi}}, \bibinfo {author} {\bibfnamefont {Peter}\ \bibnamefont
  {Jung}}, \ and\ \bibinfo {author} {\bibfnamefont {Fabio}\ \bibnamefont
  {Marchesoni}},\ }\bibfield  {title} {\enquote {\bibinfo {title} {Stochastic
  resonance},}\ }\href {\doibase 10.1103/RevModPhys.70.223} {\bibfield
  {journal} {\bibinfo  {journal} {Reviews of Modern Physics}\ }\textbf
  {\bibinfo {volume} {70}},\ \bibinfo {pages} {223} (\bibinfo {year}
  {1998})}\BibitemShut {NoStop}%
\bibitem [{\citenamefont {Lapidus}\ \emph {et~al.}(1999)\citenamefont
  {Lapidus}, \citenamefont {Enzer},\ and\ \citenamefont
  {Gabrielse}}]{lapidus1999stochastic}%
  \BibitemOpen
  \bibfield  {author} {\bibinfo {author} {\bibfnamefont {LJ}~\bibnamefont
  {Lapidus}}, \bibinfo {author} {\bibfnamefont {D}~\bibnamefont {Enzer}}, \
  and\ \bibinfo {author} {\bibfnamefont {G}~\bibnamefont {Gabrielse}},\
  }\bibfield  {title} {\enquote {\bibinfo {title} {Stochastic phase switching
  of a parametrically driven electron in a {P}enning trap},}\ }\href {\doibase
  10.1103/PhysRevLett.83.899} {\bibfield  {journal} {\bibinfo  {journal}
  {Physical Review Letters}\ }\textbf {\bibinfo {volume} {83}},\ \bibinfo
  {pages} {899} (\bibinfo {year} {1999})}\BibitemShut {NoStop}%
\bibitem [{\citenamefont {Hales}\ \emph {et~al.}(2000)\citenamefont {Hales},
  \citenamefont {Zhukov}, \citenamefont {Roy},\ and\ \citenamefont
  {Dykman}}]{hales2000dynamics}%
  \BibitemOpen
  \bibfield  {author} {\bibinfo {author} {\bibfnamefont {J}~\bibnamefont
  {Hales}}, \bibinfo {author} {\bibfnamefont {A}~\bibnamefont {Zhukov}},
  \bibinfo {author} {\bibfnamefont {R}~\bibnamefont {Roy}}, \ and\ \bibinfo
  {author} {\bibfnamefont {MI}~\bibnamefont {Dykman}},\ }\bibfield  {title}
  {\enquote {\bibinfo {title} {Dynamics of activated escape and its observation
  in a semiconductor laser},}\ }\href {\doibase 10.1103/PhysRevLett.85.78}
  {\bibfield  {journal} {\bibinfo  {journal} {Physical Review Letters}\
  }\textbf {\bibinfo {volume} {85}},\ \bibinfo {pages} {78} (\bibinfo {year}
  {2000})}\BibitemShut {NoStop}%
\bibitem [{\citenamefont {Tretiakov}\ and\ \citenamefont
  {Matveev}(2005)}]{tretiakov2005stochastic}%
  \BibitemOpen
  \bibfield  {author} {\bibinfo {author} {\bibfnamefont {OA}~\bibnamefont
  {Tretiakov}}\ and\ \bibinfo {author} {\bibfnamefont {KA}~\bibnamefont
  {Matveev}},\ }\bibfield  {title} {\enquote {\bibinfo {title} {Stochastic
  current switching in bistable resonant tunneling systems},}\ }\href {\doibase
  10.1103/PhysRevB.71.165326} {\bibfield  {journal} {\bibinfo  {journal}
  {Physical Review B}\ }\textbf {\bibinfo {volume} {71}},\ \bibinfo {pages}
  {165326} (\bibinfo {year} {2005})}\BibitemShut {NoStop}%
\bibitem [{\citenamefont {Wilkowski}\ \emph {et~al.}(2000)\citenamefont
  {Wilkowski}, \citenamefont {Ringot}, \citenamefont {Hennequin},\ and\
  \citenamefont {Garreau}}]{wilkowski2000instabilities}%
  \BibitemOpen
  \bibfield  {author} {\bibinfo {author} {\bibfnamefont {David}\ \bibnamefont
  {Wilkowski}}, \bibinfo {author} {\bibfnamefont {Jean}\ \bibnamefont
  {Ringot}}, \bibinfo {author} {\bibfnamefont {Daniel}\ \bibnamefont
  {Hennequin}}, \ and\ \bibinfo {author} {\bibfnamefont {Jean~Claude}\
  \bibnamefont {Garreau}},\ }\bibfield  {title} {\enquote {\bibinfo {title}
  {Instabilities in a magneto-optical trap: noise-induced dynamics in an atomic
  system},}\ }\href {\doibase 10.1103/PhysRevLett.85.1839} {\bibfield
  {journal} {\bibinfo  {journal} {Physical Review Letters}\ }\textbf {\bibinfo
  {volume} {85}},\ \bibinfo {pages} {1839} (\bibinfo {year}
  {2000})}\BibitemShut {NoStop}%
\bibitem [{\citenamefont {Ricci}\ \emph {et~al.}(2017)\citenamefont {Ricci},
  \citenamefont {Rica}, \citenamefont {Spasenovi{\'c}}, \citenamefont
  {Gieseler}, \citenamefont {Rondin}, \citenamefont {Novotny},\ and\
  \citenamefont {Quidant}}]{ricci2017optically}%
  \BibitemOpen
  \bibfield  {author} {\bibinfo {author} {\bibfnamefont {Francesco}\
  \bibnamefont {Ricci}}, \bibinfo {author} {\bibfnamefont {Ra{\'u}l~A}\
  \bibnamefont {Rica}}, \bibinfo {author} {\bibfnamefont {Marko}\ \bibnamefont
  {Spasenovi{\'c}}}, \bibinfo {author} {\bibfnamefont {Jan}\ \bibnamefont
  {Gieseler}}, \bibinfo {author} {\bibfnamefont {Lo{\"\i}c}\ \bibnamefont
  {Rondin}}, \bibinfo {author} {\bibfnamefont {Lukas}\ \bibnamefont {Novotny}},
  \ and\ \bibinfo {author} {\bibfnamefont {Romain}\ \bibnamefont {Quidant}},\
  }\bibfield  {title} {\enquote {\bibinfo {title} {Optically levitated
  nanoparticle as a model system for stochastic bistable dynamics},}\ }\href
  {\doibase 10.1038/ncomms15141} {\bibfield  {journal} {\bibinfo  {journal}
  {Nature Communications}\ }\textbf {\bibinfo {volume} {8}},\ \bibinfo {pages}
  {15141} (\bibinfo {year} {2017})}\BibitemShut {NoStop}%
\bibitem [{\citenamefont {Rondin}\ \emph {et~al.}(2017)\citenamefont {Rondin},
  \citenamefont {Gieseler}, \citenamefont {Ricci}, \citenamefont {Quidant},
  \citenamefont {Dellago},\ and\ \citenamefont {Novotny}}]{rondin2017direct}%
  \BibitemOpen
  \bibfield  {author} {\bibinfo {author} {\bibfnamefont {Lo{\"\i}c}\
  \bibnamefont {Rondin}}, \bibinfo {author} {\bibfnamefont {Jan}\ \bibnamefont
  {Gieseler}}, \bibinfo {author} {\bibfnamefont {Francesco}\ \bibnamefont
  {Ricci}}, \bibinfo {author} {\bibfnamefont {Romain}\ \bibnamefont {Quidant}},
  \bibinfo {author} {\bibfnamefont {Christoph}\ \bibnamefont {Dellago}}, \ and\
  \bibinfo {author} {\bibfnamefont {Lukas}\ \bibnamefont {Novotny}},\
  }\bibfield  {title} {\enquote {\bibinfo {title} {Direct measurement of
  {K}ramers turnover with a levitated nanoparticle},}\ }\href {\doibase
  10.1038/nnano.2017.198} {\bibfield  {journal} {\bibinfo  {journal} {Nature
  Nanotechnology}\ }\textbf {\bibinfo {volume} {12}},\ \bibinfo {pages} {1130}
  (\bibinfo {year} {2017})}\BibitemShut {NoStop}%
\bibitem [{\citenamefont {Levin}\ and\ \citenamefont
  {Miller}(1996)}]{levin1996broadband}%
  \BibitemOpen
  \bibfield  {author} {\bibinfo {author} {\bibfnamefont {Jacob~E}\ \bibnamefont
  {Levin}}\ and\ \bibinfo {author} {\bibfnamefont {John~P}\ \bibnamefont
  {Miller}},\ }\bibfield  {title} {\enquote {\bibinfo {title} {Broadband neural
  encoding in the cricket cereal sensory system enhanced by stochastic
  resonance},}\ }\href@noop {} {\bibfield  {journal} {\bibinfo  {journal}
  {Nature}\ }\textbf {\bibinfo {volume} {380}},\ \bibinfo {pages} {165}
  (\bibinfo {year} {1996})}\BibitemShut {NoStop}%
\bibitem [{\citenamefont {Douglass}\ \emph {et~al.}(1993)\citenamefont
  {Douglass}, \citenamefont {Wilkens}, \citenamefont {Pantazelou},\ and\
  \citenamefont {Moss}}]{douglass1993noise}%
  \BibitemOpen
  \bibfield  {author} {\bibinfo {author} {\bibfnamefont {John~K}\ \bibnamefont
  {Douglass}}, \bibinfo {author} {\bibfnamefont {Lon}\ \bibnamefont {Wilkens}},
  \bibinfo {author} {\bibfnamefont {Eleni}\ \bibnamefont {Pantazelou}}, \ and\
  \bibinfo {author} {\bibfnamefont {Frank}\ \bibnamefont {Moss}},\ }\bibfield
  {title} {\enquote {\bibinfo {title} {Noise enhancement of information
  transfer in crayfish mechanoreceptors by stochastic resonance},}\ }\href@noop
  {} {\bibfield  {journal} {\bibinfo  {journal} {Nature}\ }\textbf {\bibinfo
  {volume} {365}},\ \bibinfo {pages} {337} (\bibinfo {year}
  {1993})}\BibitemShut {NoStop}%
\bibitem [{\citenamefont {Stambaugh}\ and\ \citenamefont
  {Chan}(2006)}]{stambaugh2006noise}%
  \BibitemOpen
  \bibfield  {author} {\bibinfo {author} {\bibfnamefont {Corey}\ \bibnamefont
  {Stambaugh}}\ and\ \bibinfo {author} {\bibfnamefont {Ho~Bun}\ \bibnamefont
  {Chan}},\ }\bibfield  {title} {\enquote {\bibinfo {title} {Noise-activated
  switching in a driven nonlinear micromechanical oscillator},}\ }\href
  {\doibase 10.1103/PhysRevB.73.172302} {\bibfield  {journal} {\bibinfo
  {journal} {Physical Review B}\ }\textbf {\bibinfo {volume} {73}},\ \bibinfo
  {pages} {172302} (\bibinfo {year} {2006})}\BibitemShut {NoStop}%
\bibitem [{\citenamefont {Dykman}\ \emph {et~al.}(1998)\citenamefont {Dykman},
  \citenamefont {Maloney}, \citenamefont {Smelyanskiy},\ and\ \citenamefont
  {Silverstein}}]{dykman1998fluctuational}%
  \BibitemOpen
  \bibfield  {author} {\bibinfo {author} {\bibfnamefont {MI}~\bibnamefont
  {Dykman}}, \bibinfo {author} {\bibfnamefont {CM}~\bibnamefont {Maloney}},
  \bibinfo {author} {\bibfnamefont {VN}~\bibnamefont {Smelyanskiy}}, \ and\
  \bibinfo {author} {\bibfnamefont {M}~\bibnamefont {Silverstein}},\ }\bibfield
   {title} {\enquote {\bibinfo {title} {Fluctuational phase-flip transitions in
  parametrically driven oscillators},}\ }\href {\doibase
  10.1103/PhysRevE.57.5202} {\bibfield  {journal} {\bibinfo  {journal}
  {Physical Review E}\ }\textbf {\bibinfo {volume} {57}},\ \bibinfo {pages}
  {5202} (\bibinfo {year} {1998})}\BibitemShut {NoStop}%
\bibitem [{\citenamefont {Chan}\ \emph {et~al.}(2008)\citenamefont {Chan},
  \citenamefont {Dykman},\ and\ \citenamefont {Stambaugh}}]{chan2008paths}%
  \BibitemOpen
  \bibfield  {author} {\bibinfo {author} {\bibfnamefont {HB}~\bibnamefont
  {Chan}}, \bibinfo {author} {\bibfnamefont {Mark~I}\ \bibnamefont {Dykman}}, \
  and\ \bibinfo {author} {\bibfnamefont {Corey}\ \bibnamefont {Stambaugh}},\
  }\bibfield  {title} {\enquote {\bibinfo {title} {Paths of fluctuation induced
  switching},}\ }\href {\doibase 10.1103/PhysRevLett.100.130602} {\bibfield
  {journal} {\bibinfo  {journal} {Physical Review Letters}\ }\textbf {\bibinfo
  {volume} {100}},\ \bibinfo {pages} {130602} (\bibinfo {year}
  {2008})}\BibitemShut {NoStop}%
\bibitem [{\citenamefont {Badzey}\ and\ \citenamefont
  {Mohanty}(2005)}]{badzey2005coherent}%
  \BibitemOpen
  \bibfield  {author} {\bibinfo {author} {\bibfnamefont {Robert~L}\
  \bibnamefont {Badzey}}\ and\ \bibinfo {author} {\bibfnamefont {Pritiraj}\
  \bibnamefont {Mohanty}},\ }\bibfield  {title} {\enquote {\bibinfo {title}
  {Coherent signal amplification in bistable nanomechanical oscillators by
  stochastic resonance},}\ }\href {\doibase 10.1038/nature04124} {\bibfield
  {journal} {\bibinfo  {journal} {Nature}\ }\textbf {\bibinfo {volume} {437}},\
  \bibinfo {pages} {995} (\bibinfo {year} {2005})}\BibitemShut {NoStop}%
\bibitem [{\citenamefont {Aldridge}\ and\ \citenamefont
  {Cleland}(2005)}]{aldridge2005noise}%
  \BibitemOpen
  \bibfield  {author} {\bibinfo {author} {\bibfnamefont {JS}~\bibnamefont
  {Aldridge}}\ and\ \bibinfo {author} {\bibfnamefont {AN}~\bibnamefont
  {Cleland}},\ }\bibfield  {title} {\enquote {\bibinfo {title} {Noise-enabled
  precision measurements of a {D}uffing nanomechanical resonator},}\ }\href
  {\doibase 10.1103/PhysRevLett.94.156403} {\bibfield  {journal} {\bibinfo
  {journal} {Physical Review Letters}\ }\textbf {\bibinfo {volume} {94}},\
  \bibinfo {pages} {156403} (\bibinfo {year} {2005})}\BibitemShut {NoStop}%
\bibitem [{\citenamefont {Chan}\ and\ \citenamefont
  {Stambaugh}(2006)}]{chan2006fluctuation}%
  \BibitemOpen
  \bibfield  {author} {\bibinfo {author} {\bibfnamefont {HB}~\bibnamefont
  {Chan}}\ and\ \bibinfo {author} {\bibfnamefont {C}~\bibnamefont
  {Stambaugh}},\ }\bibfield  {title} {\enquote {\bibinfo {title}
  {Fluctuation-enhanced frequency mixing in a nonlinear micromechanical
  oscillator},}\ }\href {\doibase 10.1103/PhysRevB.73.224301} {\bibfield
  {journal} {\bibinfo  {journal} {Physical Review B}\ }\textbf {\bibinfo
  {volume} {73}},\ \bibinfo {pages} {224301} (\bibinfo {year}
  {2006})}\BibitemShut {NoStop}%
\bibitem [{\citenamefont {Ono}\ \emph {et~al.}(2008)\citenamefont {Ono},
  \citenamefont {Yoshida}, \citenamefont {Jiang},\ and\ \citenamefont
  {Esashi}}]{ono2008noise}%
  \BibitemOpen
  \bibfield  {author} {\bibinfo {author} {\bibfnamefont {Takahito}\
  \bibnamefont {Ono}}, \bibinfo {author} {\bibfnamefont {Yusuke}\ \bibnamefont
  {Yoshida}}, \bibinfo {author} {\bibfnamefont {Yong-Gang}\ \bibnamefont
  {Jiang}}, \ and\ \bibinfo {author} {\bibfnamefont {Masayoshi}\ \bibnamefont
  {Esashi}},\ }\bibfield  {title} {\enquote {\bibinfo {title} {Noise-enhanced
  sensing of light and magnetic force based on a nonlinear silicon
  microresonator},}\ }\href {\doibase 10.1143/APEX.1.123001} {\bibfield
  {journal} {\bibinfo  {journal} {Applied Physics Express}\ }\textbf {\bibinfo
  {volume} {1}},\ \bibinfo {pages} {123001} (\bibinfo {year}
  {2008})}\BibitemShut {NoStop}%
\bibitem [{\citenamefont {Venstra}\ \emph {et~al.}(2013)\citenamefont
  {Venstra}, \citenamefont {Westra},\ and\ \citenamefont {Van
  Der~Zant}}]{venstra2013stochastic}%
  \BibitemOpen
  \bibfield  {author} {\bibinfo {author} {\bibfnamefont {Warner~J}\
  \bibnamefont {Venstra}}, \bibinfo {author} {\bibfnamefont {Hidde~JR}\
  \bibnamefont {Westra}}, \ and\ \bibinfo {author} {\bibfnamefont {Herre~SJ}\
  \bibnamefont {Van Der~Zant}},\ }\bibfield  {title} {\enquote {\bibinfo
  {title} {Stochastic switching of cantilever motion},}\ }\href {\doibase
  10.1038/ncomms3624} {\bibfield  {journal} {\bibinfo  {journal} {Nature
  Communications}\ }\textbf {\bibinfo {volume} {4}},\ \bibinfo {pages} {2624}
  (\bibinfo {year} {2013})}\BibitemShut {NoStop}%
\bibitem [{\citenamefont {Novoselov}\ \emph {et~al.}(2005)\citenamefont
  {Novoselov}, \citenamefont {Jiang}, \citenamefont {Schedin}, \citenamefont
  {Booth}, \citenamefont {Khotkevich}, \citenamefont {Morozov},\ and\
  \citenamefont {Geim}}]{novoselov2005two2}%
  \BibitemOpen
  \bibfield  {author} {\bibinfo {author} {\bibfnamefont {KS}~\bibnamefont
  {Novoselov}}, \bibinfo {author} {\bibfnamefont {D}~\bibnamefont {Jiang}},
  \bibinfo {author} {\bibfnamefont {F}~\bibnamefont {Schedin}}, \bibinfo
  {author} {\bibfnamefont {TJ}~\bibnamefont {Booth}}, \bibinfo {author}
  {\bibfnamefont {VV}~\bibnamefont {Khotkevich}}, \bibinfo {author}
  {\bibfnamefont {SV}~\bibnamefont {Morozov}}, \ and\ \bibinfo {author}
  {\bibfnamefont {AK}~\bibnamefont {Geim}},\ }\bibfield  {title} {\enquote
  {\bibinfo {title} {Two-dimensional atomic crystals},}\ }\href {\doibase
  10.1073/pnas.0502848102} {\bibfield  {journal} {\bibinfo  {journal}
  {Proceedings of the National Academy of Sciences}\ }\textbf {\bibinfo
  {volume} {102}},\ \bibinfo {pages} {10451--10453} (\bibinfo {year}
  {2005})}\BibitemShut {NoStop}%
\bibitem [{\citenamefont {Geim}\ and\ \citenamefont
  {Novoselov}(2007)}]{geim2007rise}%
  \BibitemOpen
  \bibfield  {author} {\bibinfo {author} {\bibfnamefont {Andre~K}\ \bibnamefont
  {Geim}}\ and\ \bibinfo {author} {\bibfnamefont {Konstantin~S}\ \bibnamefont
  {Novoselov}},\ }\bibfield  {title} {\enquote {\bibinfo {title} {The rise of
  graphene},}\ }\href {\doibase 10.1038/nmat1849} {\bibfield  {journal}
  {\bibinfo  {journal} {Nature Materials}\ }\textbf {\bibinfo {volume} {6}},\
  \bibinfo {pages} {183--191} (\bibinfo {year} {2007})}\BibitemShut {NoStop}%
\bibitem [{\citenamefont {Lee}\ \emph {et~al.}(2008)\citenamefont {Lee},
  \citenamefont {Wei}, \citenamefont {Kysar},\ and\ \citenamefont
  {Hone}}]{lee2008measurement}%
  \BibitemOpen
  \bibfield  {author} {\bibinfo {author} {\bibfnamefont {Changgu}\ \bibnamefont
  {Lee}}, \bibinfo {author} {\bibfnamefont {Xiaoding}\ \bibnamefont {Wei}},
  \bibinfo {author} {\bibfnamefont {Jeffrey~W.}\ \bibnamefont {Kysar}}, \ and\
  \bibinfo {author} {\bibfnamefont {James}\ \bibnamefont {Hone}},\ }\bibfield
  {title} {\enquote {\bibinfo {title} {Measurement of the elastic properties
  and intrinsic strength of monolayer graphene},}\ }\href {\doibase
  10.1126/science.1157996} {\bibfield  {journal} {\bibinfo  {journal}
  {Science}\ }\textbf {\bibinfo {volume} {321}},\ \bibinfo {pages} {385--388}
  (\bibinfo {year} {2008})}\BibitemShut {NoStop}%
\bibitem [{\citenamefont {Bunch}\ \emph {et~al.}(2007)\citenamefont {Bunch},
  \citenamefont {van~der Zande}, \citenamefont {Verbridge}, \citenamefont
  {Frank}, \citenamefont {Tanenbaum}, \citenamefont {Parpia}, \citenamefont
  {Craighead},\ and\ \citenamefont {McEuen}}]{bunch2007electromechanical}%
  \BibitemOpen
  \bibfield  {author} {\bibinfo {author} {\bibfnamefont {J.~Scott}\
  \bibnamefont {Bunch}}, \bibinfo {author} {\bibfnamefont {Arend~M.}\
  \bibnamefont {van~der Zande}}, \bibinfo {author} {\bibfnamefont {Scott~S.}\
  \bibnamefont {Verbridge}}, \bibinfo {author} {\bibfnamefont {Ian~W.}\
  \bibnamefont {Frank}}, \bibinfo {author} {\bibfnamefont {David~M.}\
  \bibnamefont {Tanenbaum}}, \bibinfo {author} {\bibfnamefont {Jeevak~M.}\
  \bibnamefont {Parpia}}, \bibinfo {author} {\bibfnamefont {Harold~G.}\
  \bibnamefont {Craighead}}, \ and\ \bibinfo {author} {\bibfnamefont {Paul~L.}\
  \bibnamefont {McEuen}},\ }\bibfield  {title} {\enquote {\bibinfo {title}
  {Electromechanical resonators from graphene sheets},}\ }\href {\doibase
  10.1126/science.1136836} {\bibfield  {journal} {\bibinfo  {journal}
  {Science}\ }\textbf {\bibinfo {volume} {315}},\ \bibinfo {pages} {490--493}
  (\bibinfo {year} {2007})}\BibitemShut {NoStop}%
\bibitem [{\citenamefont {Chen}\ \emph {et~al.}(2009)\citenamefont {Chen},
  \citenamefont {Rosenblatt}, \citenamefont {Bolotin}, \citenamefont {Kalb},
  \citenamefont {Kim}, \citenamefont {Kymissis}, \citenamefont {Stormer},
  \citenamefont {Heinz},\ and\ \citenamefont {Hone}}]{chen2009performance}%
  \BibitemOpen
  \bibfield  {author} {\bibinfo {author} {\bibfnamefont {Changyao}\
  \bibnamefont {Chen}}, \bibinfo {author} {\bibfnamefont {Sami}\ \bibnamefont
  {Rosenblatt}}, \bibinfo {author} {\bibfnamefont {Kirill~I}\ \bibnamefont
  {Bolotin}}, \bibinfo {author} {\bibfnamefont {William}\ \bibnamefont {Kalb}},
  \bibinfo {author} {\bibfnamefont {Philip}\ \bibnamefont {Kim}}, \bibinfo
  {author} {\bibfnamefont {Ioannis}\ \bibnamefont {Kymissis}}, \bibinfo
  {author} {\bibfnamefont {Horst~L}\ \bibnamefont {Stormer}}, \bibinfo {author}
  {\bibfnamefont {Tony~F}\ \bibnamefont {Heinz}}, \ and\ \bibinfo {author}
  {\bibfnamefont {James}\ \bibnamefont {Hone}},\ }\bibfield  {title} {\enquote
  {\bibinfo {title} {Performance of monolayer graphene nanomechanical
  resonators with electrical readout},}\ }\href {\doibase
  10.1038/nnano.2009.267} {\bibfield  {journal} {\bibinfo  {journal} {Nature
  Nanotechnology}\ }\textbf {\bibinfo {volume} {4}},\ \bibinfo {pages}
  {861--867} (\bibinfo {year} {2009})}\BibitemShut {NoStop}%
\bibitem [{\citenamefont {Davidovikj}\ \emph {et~al.}(2017)\citenamefont
  {Davidovikj}, \citenamefont {Alijani}, \citenamefont {Cartamil-Bueno},
  \citenamefont {van~der Zant}, \citenamefont {Amabili},\ and\ \citenamefont
  {Steeneken}}]{davidovikj2017nonlinear}%
  \BibitemOpen
  \bibfield  {author} {\bibinfo {author} {\bibfnamefont {D.}~\bibnamefont
  {Davidovikj}}, \bibinfo {author} {\bibfnamefont {F.}~\bibnamefont {Alijani}},
  \bibinfo {author} {\bibfnamefont {S.~J.}\ \bibnamefont {Cartamil-Bueno}},
  \bibinfo {author} {\bibfnamefont {H.~S.~J.}\ \bibnamefont {van~der Zant}},
  \bibinfo {author} {\bibfnamefont {M.}~\bibnamefont {Amabili}}, \ and\
  \bibinfo {author} {\bibfnamefont {P.~G.}\ \bibnamefont {Steeneken}},\
  }\bibfield  {title} {\enquote {\bibinfo {title} {Nonlinear dynamic
  characterization of two-dimensional materials},}\ }\href {\doibase
  10.1038/s41467-017-01351-4} {\bibfield  {journal} {\bibinfo  {journal}
  {Nature Communications}\ }\textbf {\bibinfo {volume} {8}},\ \bibinfo {pages}
  {1253} (\bibinfo {year} {2017})}\BibitemShut {NoStop}%
\bibitem [{\citenamefont {Dolleman}\ \emph {et~al.}(2018)\citenamefont
  {Dolleman}, \citenamefont {Houri}, \citenamefont {Chandrashekar},
  \citenamefont {Alijani}, \citenamefont {van~der Zant},\ and\ \citenamefont
  {Steeneken}}]{dolleman2018opto}%
  \BibitemOpen
  \bibfield  {author} {\bibinfo {author} {\bibfnamefont {Robin~J}\ \bibnamefont
  {Dolleman}}, \bibinfo {author} {\bibfnamefont {Samer}\ \bibnamefont {Houri}},
  \bibinfo {author} {\bibfnamefont {Abhilash}\ \bibnamefont {Chandrashekar}},
  \bibinfo {author} {\bibfnamefont {Farbod}\ \bibnamefont {Alijani}}, \bibinfo
  {author} {\bibfnamefont {Herre~SJ}\ \bibnamefont {van~der Zant}}, \ and\
  \bibinfo {author} {\bibfnamefont {Peter~G}\ \bibnamefont {Steeneken}},\
  }\bibfield  {title} {\enquote {\bibinfo {title} {Opto-thermally excited
  multimode parametric resonance in graphene membranes},}\ }\href {\doibase
  10.1038/s41598-018-27561-4} {\bibfield  {journal} {\bibinfo  {journal}
  {Scientific Reports}\ }\textbf {\bibinfo {volume} {8}},\ \bibinfo {pages}
  {9366} (\bibinfo {year} {2018})}\BibitemShut {NoStop}%
\bibitem [{\citenamefont {Bunch}\ \emph {et~al.}(2008)\citenamefont {Bunch},
  \citenamefont {Verbridge}, \citenamefont {Alden}, \citenamefont {Van
  Der~Zande}, \citenamefont {Parpia}, \citenamefont {Craighead},\ and\
  \citenamefont {McEuen}}]{bunch2008impermeable}%
  \BibitemOpen
  \bibfield  {author} {\bibinfo {author} {\bibfnamefont {J~Scott}\ \bibnamefont
  {Bunch}}, \bibinfo {author} {\bibfnamefont {Scott~S}\ \bibnamefont
  {Verbridge}}, \bibinfo {author} {\bibfnamefont {Jonathan~S}\ \bibnamefont
  {Alden}}, \bibinfo {author} {\bibfnamefont {Arend~M}\ \bibnamefont {Van
  Der~Zande}}, \bibinfo {author} {\bibfnamefont {Jeevak~M}\ \bibnamefont
  {Parpia}}, \bibinfo {author} {\bibfnamefont {Harold~G}\ \bibnamefont
  {Craighead}}, \ and\ \bibinfo {author} {\bibfnamefont {Paul~L}\ \bibnamefont
  {McEuen}},\ }\bibfield  {title} {\enquote {\bibinfo {title} {Impermeable
  atomic membranes from graphene sheets},}\ }\href {\doibase 10.1021/nl801457b}
  {\bibfield  {journal} {\bibinfo  {journal} {Nano Letters}\ }\textbf {\bibinfo
  {volume} {8}},\ \bibinfo {pages} {2458--2462} (\bibinfo {year}
  {2008})}\BibitemShut {NoStop}%
\bibitem [{\citenamefont {Smith}\ \emph {et~al.}(2013)\citenamefont {Smith},
  \citenamefont {Niklaus}, \citenamefont {Paussa}, \citenamefont {Vaziri},
  \citenamefont {Fischer}, \citenamefont {Sterner}, \citenamefont {Forsberg},
  \citenamefont {Delin}, \citenamefont {Esseni}, \citenamefont {Palestri},
  \citenamefont {Ostling},\ and\ \citenamefont
  {Lemme}}]{smith2013electromechanical}%
  \BibitemOpen
  \bibfield  {author} {\bibinfo {author} {\bibfnamefont {AD}~\bibnamefont
  {Smith}}, \bibinfo {author} {\bibfnamefont {Frank}\ \bibnamefont {Niklaus}},
  \bibinfo {author} {\bibfnamefont {A}~\bibnamefont {Paussa}}, \bibinfo
  {author} {\bibfnamefont {Sam}\ \bibnamefont {Vaziri}}, \bibinfo {author}
  {\bibfnamefont {Andreas~C}\ \bibnamefont {Fischer}}, \bibinfo {author}
  {\bibfnamefont {Mikael}\ \bibnamefont {Sterner}}, \bibinfo {author}
  {\bibfnamefont {Fredrik}\ \bibnamefont {Forsberg}}, \bibinfo {author}
  {\bibfnamefont {Anna}\ \bibnamefont {Delin}}, \bibinfo {author}
  {\bibfnamefont {D}~\bibnamefont {Esseni}}, \bibinfo {author} {\bibfnamefont
  {P}~\bibnamefont {Palestri}}, \bibinfo {author} {\bibfnamefont
  {M}~\bibnamefont {Ostling}}, \ and\ \bibinfo {author} {\bibfnamefont
  {MC}~\bibnamefont {Lemme}},\ }\bibfield  {title} {\enquote {\bibinfo {title}
  {Electromechanical piezoresistive sensing in suspended graphene membranes},}\
  }\href {\doibase 10.1021/nl401352k} {\bibfield  {journal} {\bibinfo
  {journal} {Nano Letters}\ }\textbf {\bibinfo {volume} {13}},\ \bibinfo
  {pages} {3237--3242} (\bibinfo {year} {2013})}\BibitemShut {NoStop}%
\bibitem [{\citenamefont {Dolleman}\ \emph {et~al.}(2016)\citenamefont
  {Dolleman}, \citenamefont {Davidovikj}, \citenamefont {Cartamil-Bueno},
  \citenamefont {van~der Zant},\ and\ \citenamefont
  {Steeneken}}]{dolleman2015graphene}%
  \BibitemOpen
  \bibfield  {author} {\bibinfo {author} {\bibfnamefont {Robin~J}\ \bibnamefont
  {Dolleman}}, \bibinfo {author} {\bibfnamefont {Dejan}\ \bibnamefont
  {Davidovikj}}, \bibinfo {author} {\bibfnamefont {Santiago~J}\ \bibnamefont
  {Cartamil-Bueno}}, \bibinfo {author} {\bibfnamefont {Herre~SJ}\ \bibnamefont
  {van~der Zant}}, \ and\ \bibinfo {author} {\bibfnamefont {Peter~G}\
  \bibnamefont {Steeneken}},\ }\bibfield  {title} {\enquote {\bibinfo {title}
  {Graphene squeeze-film pressure sensors},}\ }\href {\doibase
  10.1021/acs.nanolett.5b04251} {\bibfield  {journal} {\bibinfo  {journal}
  {Nano Letters}\ }\textbf {\bibinfo {volume} {16}},\ \bibinfo {pages}
  {568--571} (\bibinfo {year} {2016})}\BibitemShut {NoStop}%
\bibitem [{\citenamefont {Eichler}\ \emph {et~al.}(2011)\citenamefont
  {Eichler}, \citenamefont {Moser}, \citenamefont {Chaste}, \citenamefont
  {Zdrojek}, \citenamefont {Wilson-Rae},\ and\ \citenamefont
  {Bachtold}}]{eichler2011nonlinear}%
  \BibitemOpen
  \bibfield  {author} {\bibinfo {author} {\bibfnamefont {A}~\bibnamefont
  {Eichler}}, \bibinfo {author} {\bibfnamefont {Joel}\ \bibnamefont {Moser}},
  \bibinfo {author} {\bibfnamefont {J}~\bibnamefont {Chaste}}, \bibinfo
  {author} {\bibfnamefont {M}~\bibnamefont {Zdrojek}}, \bibinfo {author}
  {\bibfnamefont {I}~\bibnamefont {Wilson-Rae}}, \ and\ \bibinfo {author}
  {\bibfnamefont {Adrian}\ \bibnamefont {Bachtold}},\ }\bibfield  {title}
  {\enquote {\bibinfo {title} {Nonlinear damping in mechanical resonators made
  from carbon nanotubes and graphene},}\ }\href {\doibase
  10.1038/nnano.2011.71} {\bibfield  {journal} {\bibinfo  {journal} {Nature
  Nanotechnology}\ }\textbf {\bibinfo {volume} {6}},\ \bibinfo {pages}
  {339--342} (\bibinfo {year} {2011})}\BibitemShut {NoStop}%
\bibitem [{\citenamefont {Koenig}\ \emph {et~al.}(2012)\citenamefont {Koenig},
  \citenamefont {Wang}, \citenamefont {Pellegrino},\ and\ \citenamefont
  {Bunch}}]{koenig2012selective}%
  \BibitemOpen
  \bibfield  {author} {\bibinfo {author} {\bibfnamefont {Steven~P}\
  \bibnamefont {Koenig}}, \bibinfo {author} {\bibfnamefont {Luda}\ \bibnamefont
  {Wang}}, \bibinfo {author} {\bibfnamefont {John}\ \bibnamefont {Pellegrino}},
  \ and\ \bibinfo {author} {\bibfnamefont {J~Scott}\ \bibnamefont {Bunch}},\
  }\bibfield  {title} {\enquote {\bibinfo {title} {Selective molecular sieving
  through porous graphene},}\ }\href {\doibase 10.1038/nnano.2012.162}
  {\bibfield  {journal} {\bibinfo  {journal} {Nature Nanotechnology}\ }\textbf
  {\bibinfo {volume} {7}},\ \bibinfo {pages} {728--732} (\bibinfo {year}
  {2012})}\BibitemShut {NoStop}%
\bibitem [{\citenamefont {Dolleman}\ \emph
  {et~al.}(2017{\natexlab{a}})\citenamefont {Dolleman}, \citenamefont
  {Cartamil-Bueno}, \citenamefont {van~der Zant},\ and\ \citenamefont
  {Steeneken}}]{dolleman2016graphene}%
  \BibitemOpen
  \bibfield  {author} {\bibinfo {author} {\bibfnamefont {Robin~J}\ \bibnamefont
  {Dolleman}}, \bibinfo {author} {\bibfnamefont {Santiago~J}\ \bibnamefont
  {Cartamil-Bueno}}, \bibinfo {author} {\bibfnamefont {Herre S~J}\ \bibnamefont
  {van~der Zant}}, \ and\ \bibinfo {author} {\bibfnamefont {Peter~G}\
  \bibnamefont {Steeneken}},\ }\bibfield  {title} {\enquote {\bibinfo {title}
  {Graphene gas osmometers},}\ }\href {\doibase 10.1088/2053-1583/4/1/011002}
  {\bibfield  {journal} {\bibinfo  {journal} {2D Materials}\ }\textbf {\bibinfo
  {volume} {4}},\ \bibinfo {pages} {011002} (\bibinfo {year}
  {2017}{\natexlab{a}})}\BibitemShut {NoStop}%
\bibitem [{\citenamefont {Kramers}(1940)}]{kramers1940brownian}%
  \BibitemOpen
  \bibfield  {author} {\bibinfo {author} {\bibfnamefont {Hendrik~Anthony}\
  \bibnamefont {Kramers}},\ }\bibfield  {title} {\enquote {\bibinfo {title}
  {Brownian motion in a field of force and the diffusion model of chemical
  reactions},}\ }\href@noop {} {\bibfield  {journal} {\bibinfo  {journal}
  {Physica}\ }\textbf {\bibinfo {volume} {7}},\ \bibinfo {pages} {284--304}
  (\bibinfo {year} {1940})}\BibitemShut {NoStop}%
\bibitem [{\citenamefont {Dolleman}\ \emph
  {et~al.}(2017{\natexlab{b}})\citenamefont {Dolleman}, \citenamefont {Houri},
  \citenamefont {Davidovikj}, \citenamefont {Cartamil-Bueno}, \citenamefont
  {Blanter}, \citenamefont {van~der Zant},\ and\ \citenamefont
  {Steeneken}}]{dolleman2017optomechanics}%
  \BibitemOpen
  \bibfield  {author} {\bibinfo {author} {\bibfnamefont {Robin~J}\ \bibnamefont
  {Dolleman}}, \bibinfo {author} {\bibfnamefont {Samer}\ \bibnamefont {Houri}},
  \bibinfo {author} {\bibfnamefont {Dejan}\ \bibnamefont {Davidovikj}},
  \bibinfo {author} {\bibfnamefont {Santiago~J}\ \bibnamefont
  {Cartamil-Bueno}}, \bibinfo {author} {\bibfnamefont {Yaroslav~M}\
  \bibnamefont {Blanter}}, \bibinfo {author} {\bibfnamefont {Herre~SJ}\
  \bibnamefont {van~der Zant}}, \ and\ \bibinfo {author} {\bibfnamefont
  {Peter~G}\ \bibnamefont {Steeneken}},\ }\bibfield  {title} {\enquote
  {\bibinfo {title} {Optomechanics for thermal characterization of suspended
  graphene},}\ }\href {\doibase 10.1103/PhysRevB.96.165421} {\bibfield
  {journal} {\bibinfo  {journal} {Physical Review B}\ }\textbf {\bibinfo
  {volume} {96}},\ \bibinfo {pages} {165421} (\bibinfo {year}
  {2017}{\natexlab{b}})}\BibitemShut {NoStop}%
\bibitem [{\citenamefont {Dolleman}\ \emph
  {et~al.}(2017{\natexlab{c}})\citenamefont {Dolleman}, \citenamefont
  {Davidovikj}, \citenamefont {van~der Zant},\ and\ \citenamefont
  {Steeneken}}]{dolleman2017amplitude}%
  \BibitemOpen
  \bibfield  {author} {\bibinfo {author} {\bibfnamefont {Robin~J.}\
  \bibnamefont {Dolleman}}, \bibinfo {author} {\bibfnamefont {Dejan}\
  \bibnamefont {Davidovikj}}, \bibinfo {author} {\bibfnamefont {Herre S.~J.}\
  \bibnamefont {van~der Zant}}, \ and\ \bibinfo {author} {\bibfnamefont
  {Peter~G.}\ \bibnamefont {Steeneken}},\ }\bibfield  {title} {\enquote
  {\bibinfo {title} {Amplitude calibration of {2D} mechanical resonators by
  nonlinear optical transduction},}\ }\href {\doibase 10.1063/1.5009909}
  {\bibfield  {journal} {\bibinfo  {journal} {Applied Physics Letters}\
  }\textbf {\bibinfo {volume} {111}},\ \bibinfo {pages} {253104} (\bibinfo
  {year} {2017}{\natexlab{c}})}\BibitemShut {NoStop}%
\bibitem [{\citenamefont {Hauer}\ \emph {et~al.}(2013)\citenamefont {Hauer},
  \citenamefont {Doolin}, \citenamefont {Beach},\ and\ \citenamefont
  {Davis}}]{hauer2013general}%
  \BibitemOpen
  \bibfield  {author} {\bibinfo {author} {\bibfnamefont {BD}~\bibnamefont
  {Hauer}}, \bibinfo {author} {\bibfnamefont {C}~\bibnamefont {Doolin}},
  \bibinfo {author} {\bibfnamefont {KSD}\ \bibnamefont {Beach}}, \ and\
  \bibinfo {author} {\bibfnamefont {JP}~\bibnamefont {Davis}},\ }\bibfield
  {title} {\enquote {\bibinfo {title} {A general procedure for thermomechanical
  calibration of nano/micro-mechanical resonators},}\ }\href {\doibase
  10.1016/j.aop.2013.08.003} {\bibfield  {journal} {\bibinfo  {journal} {Annals
  of Physics}\ }\textbf {\bibinfo {volume} {339}},\ \bibinfo {pages} {181--207}
  (\bibinfo {year} {2013})}\BibitemShut {NoStop}%
\bibitem [{\citenamefont {Kryloff}\ \emph {et~al.}(1947)\citenamefont
  {Kryloff}, \citenamefont {Bogoliubov},\ and\ \citenamefont
  {Lefschetz}}]{kryloff1947introduction}%
  \BibitemOpen
  \bibfield  {author} {\bibinfo {author} {\bibfnamefont {N.}~\bibnamefont
  {Kryloff}}, \bibinfo {author} {\bibfnamefont {N.N.}\ \bibnamefont
  {Bogoliubov}}, \ and\ \bibinfo {author} {\bibfnamefont {S.}~\bibnamefont
  {Lefschetz}},\ }\href@noop {} {\emph {\bibinfo {title} {Introduction to
  Non-linear Mechanics}}},\ Annals of mathematics studies\ (\bibinfo
  {publisher} {Princeton University Press},\ \bibinfo {year}
  {1947})\BibitemShut {NoStop}%
\bibitem [{\citenamefont {Zhou}\ \emph {et~al.}(2012)\citenamefont {Zhou},
  \citenamefont {Aliyu}, \citenamefont {Aurell},\ and\ \citenamefont
  {Huang}}]{Zhou2012}%
  \BibitemOpen
  \bibfield  {author} {\bibinfo {author} {\bibfnamefont {Joseph~Xu}\
  \bibnamefont {Zhou}}, \bibinfo {author} {\bibfnamefont {M~D~S}\ \bibnamefont
  {Aliyu}}, \bibinfo {author} {\bibfnamefont {Erik}\ \bibnamefont {Aurell}}, \
  and\ \bibinfo {author} {\bibfnamefont {Sui}\ \bibnamefont {Huang}},\
  }\bibfield  {title} {\enquote {\bibinfo {title} {{Quasi-potential landscape
  in complex multi-stable systems}},}\ }\href {\doibase 10.1098/rsif.2012.0434}
  {\bibfield  {journal} {\bibinfo  {journal} {Journal of the Royal Society,
  Interface}\ }\textbf {\bibinfo {volume} {9}},\ \bibinfo {pages} {3539--3553}
  (\bibinfo {year} {2012})}\BibitemShut {NoStop}%
\bibitem [{\citenamefont {Nolting}\ and\ \citenamefont
  {Abbott}(2016)}]{Nolting2015}%
  \BibitemOpen
  \bibfield  {author} {\bibinfo {author} {\bibfnamefont {Ben~C.}\ \bibnamefont
  {Nolting}}\ and\ \bibinfo {author} {\bibfnamefont {Karen~C.}\ \bibnamefont
  {Abbott}},\ }\bibfield  {title} {\enquote {\bibinfo {title} {Balls, cups, and
  quasi-potentials: quantifying stability in stochastic systems},}\ }\href
  {\doibase 10.1890/15-1047.1} {\bibfield  {journal} {\bibinfo  {journal}
  {Ecology}\ }\textbf {\bibinfo {volume} {97}},\ \bibinfo {pages} {850--864}
  (\bibinfo {year} {2016})},\ \Eprint
  {http://arxiv.org/abs/https://esajournals.onlinelibrary.wiley.com/doi/pdf/10.1890/15-1047.1}
  {https://esajournals.onlinelibrary.wiley.com/doi/pdf/10.1890/15-1047.1}
  \BibitemShut {NoStop}%
\bibitem [{\citenamefont {Sethian}\ and\ \citenamefont
  {Vladimirsky}(2001)}]{Sethian11069}%
  \BibitemOpen
  \bibfield  {author} {\bibinfo {author} {\bibfnamefont {J.~A.}\ \bibnamefont
  {Sethian}}\ and\ \bibinfo {author} {\bibfnamefont {A.}~\bibnamefont
  {Vladimirsky}},\ }\bibfield  {title} {\enquote {\bibinfo {title} {Ordered
  upwind methods for static hamilton{\textendash}jacobi equations},}\ }\href
  {\doibase 10.1073/pnas.201222998} {\bibfield  {journal} {\bibinfo  {journal}
  {Proceedings of the National Academy of Sciences}\ }\textbf {\bibinfo
  {volume} {98}},\ \bibinfo {pages} {11069--11074} (\bibinfo {year}
  {2001})}\BibitemShut {NoStop}%
\bibitem [{\citenamefont {Moore}\ \emph {et~al.}(2015)\citenamefont {Moore},
  \citenamefont {Stieha}, \citenamefont {Nolting}, \citenamefont {Cameron},\
  and\ \citenamefont {Abbott}}]{Moore029777}%
  \BibitemOpen
  \bibfield  {author} {\bibinfo {author} {\bibfnamefont {Christopher~M.}\
  \bibnamefont {Moore}}, \bibinfo {author} {\bibfnamefont {Christopher~R.}\
  \bibnamefont {Stieha}}, \bibinfo {author} {\bibfnamefont {Ben~C.}\
  \bibnamefont {Nolting}}, \bibinfo {author} {\bibfnamefont {Maria~K.}\
  \bibnamefont {Cameron}}, \ and\ \bibinfo {author} {\bibfnamefont {Karen~C.}\
  \bibnamefont {Abbott}},\ }\bibfield  {title} {\enquote {\bibinfo {title}
  {Qpot: An r package for stochastic differential equation quasi-potential
  analysis},}\ }\href {\doibase 10.1101/029777} {\bibfield  {journal} {\bibinfo
   {journal} {bioRxiv}\ } (\bibinfo {year} {2015}),\
  10.1101/029777}\BibitemShut {NoStop}%
\bibitem [{\citenamefont {Song}\ \emph {et~al.}(2011)\citenamefont {Song},
  \citenamefont {Schneider}, \citenamefont {Xu}, \citenamefont {Pandraud},
  \citenamefont {Dekker},\ and\ \citenamefont {Zandbergen}}]{song2011atomic}%
  \BibitemOpen
  \bibfield  {author} {\bibinfo {author} {\bibfnamefont {Bo}~\bibnamefont
  {Song}}, \bibinfo {author} {\bibfnamefont {Gr{\'e}gory~F}\ \bibnamefont
  {Schneider}}, \bibinfo {author} {\bibfnamefont {Qiang}\ \bibnamefont {Xu}},
  \bibinfo {author} {\bibfnamefont {Gr{\'e}gory}\ \bibnamefont {Pandraud}},
  \bibinfo {author} {\bibfnamefont {Cees}\ \bibnamefont {Dekker}}, \ and\
  \bibinfo {author} {\bibfnamefont {Henny}\ \bibnamefont {Zandbergen}},\
  }\bibfield  {title} {\enquote {\bibinfo {title} {Atomic-scale electron-beam
  sculpting of near-defect-free graphene nanostructures},}\ }\href@noop {}
  {\bibfield  {journal} {\bibinfo  {journal} {Nano letters}\ }\textbf {\bibinfo
  {volume} {11}},\ \bibinfo {pages} {2247--2250} (\bibinfo {year}
  {2011})}\BibitemShut {NoStop}%
\bibitem [{\citenamefont {Fischbein}\ and\ \citenamefont
  {Drndi{\'c}}(2008)}]{fischbein2008electron}%
  \BibitemOpen
  \bibfield  {author} {\bibinfo {author} {\bibfnamefont {Michael~D}\
  \bibnamefont {Fischbein}}\ and\ \bibinfo {author} {\bibfnamefont {Marija}\
  \bibnamefont {Drndi{\'c}}},\ }\bibfield  {title} {\enquote {\bibinfo {title}
  {Electron beam nanosculpting of suspended graphene sheets},}\ }\href@noop {}
  {\bibfield  {journal} {\bibinfo  {journal} {Applied physics letters}\
  }\textbf {\bibinfo {volume} {93}},\ \bibinfo {pages} {113107} (\bibinfo
  {year} {2008})}\BibitemShut {NoStop}%
\bibitem [{\citenamefont {Cai}\ \emph {et~al.}(2010)\citenamefont {Cai},
  \citenamefont {Moore}, \citenamefont {Zhu}, \citenamefont {Li}, \citenamefont
  {Chen}, \citenamefont {Shi},\ and\ \citenamefont {Ruoff}}]{cai2010thermal}%
  \BibitemOpen
  \bibfield  {author} {\bibinfo {author} {\bibfnamefont {Weiwei}\ \bibnamefont
  {Cai}}, \bibinfo {author} {\bibfnamefont {Arden~L}\ \bibnamefont {Moore}},
  \bibinfo {author} {\bibfnamefont {Yanwu}\ \bibnamefont {Zhu}}, \bibinfo
  {author} {\bibfnamefont {Xuesong}\ \bibnamefont {Li}}, \bibinfo {author}
  {\bibfnamefont {Shanshan}\ \bibnamefont {Chen}}, \bibinfo {author}
  {\bibfnamefont {Li}~\bibnamefont {Shi}}, \ and\ \bibinfo {author}
  {\bibfnamefont {Rodney~S}\ \bibnamefont {Ruoff}},\ }\bibfield  {title}
  {\enquote {\bibinfo {title} {Thermal transport in suspended and supported
  monolayer graphene grown by chemical vapor deposition},}\ }\href {\doibase
  10.1021/nl9041966} {\bibfield  {journal} {\bibinfo  {journal} {Nano Letters}\
  }\textbf {\bibinfo {volume} {10}},\ \bibinfo {pages} {1645--1651} (\bibinfo
  {year} {2010})}\BibitemShut {NoStop}%
\bibitem [{\citenamefont {Oksendal}(2013)}]{oksendal2013stochastic}%
  \BibitemOpen
  \bibfield  {author} {\bibinfo {author} {\bibfnamefont {B.}~\bibnamefont
  {Oksendal}},\ }\href {\doibase 10.1007/978-3-642-14394-6} {\emph {\bibinfo
  {title} {Stochastic Differential Equations: An Introduction with
  Applications}}},\ Universitext\ (\bibinfo  {publisher} {Springer Berlin
  Heidelberg},\ \bibinfo {year} {2013})\BibitemShut {NoStop}%
\end{thebibliography}
\end{document}